\documentclass[12pt,preprint]{aastex}
\usepackage{multirow}
\usepackage{booktabs}
\usepackage{graphicx}
\usepackage{amsmath}
\usepackage{longtable}
\usepackage{CJKutf8}
\usepackage[caption=false]{subfig}

\newcommand{\msun}{M_\odot}

\newcommand{\mplanet}{M_{\rm p}}

\newcommand{\mj}{M_{\rm J}}
\newcommand{\rp}{R_{\rm p}}
\newcommand{\rsub}{r_{\rm sub}}
\newcommand{\rin}{r_{\rm in,hydro}}
\newcommand{\rout}{r_{\rm out,hydro}}
\newcommand{\rhodust}{\rho_{\rm dust}}
\newcommand{\rhogas}{\rho_{\rm gas}}

\begin{document}
\begin{CJK*}{UTF8}{bsmi}
\title{How Spirals and Gaps Driven by Companions in Protoplanetary Disks Appear in Scattered Light at Arbitrary Viewing Angles}

\shortauthors{Dong, Fung, \& Chiang}

\author{Ruobing Dong (董若冰)\altaffilmark{1,2,3}, Jeffrey Fung (馮澤之)\altaffilmark{2,4}, Eugene Chiang (蔣詒曾)\altaffilmark{2}}

\altaffiltext{1}{Nuclear Science Division, Lawrence Berkeley National Lab, Berkeley, CA 94720, rdong2013@berkeley.edu}
\altaffiltext{2}{Department of Astronomy, University of California at Berkeley, Berkeley, CA 94720}
\altaffiltext{3}{NASA Hubble Fellow}
\altaffiltext{4}{NSERC Fellow}

\clearpage

\begin{abstract}

Direct imaging observations of
protoplanetary disks at near-infrared
(NIR) wavelengths 
have revealed structures 
of potentially planetary origin.
Investigations of observational signatures from planet-induced features have so far focused on disks viewed face-on. Combining 3D hydrodynamics and radiative transfer simulations, we study how the appearance of the spiral arms and the gap produced in a disk by a companion varies with inclination and position angle in NIR scattered light. We compare the cases of a $3\mj$ and a $0.1\msun$ companion, and make predictions suitable for testing with Gemini/GPI, VLT/NACO/SPHERE, and Subaru/HiCIAO/SCExAO. We find that the two trailing arms produced by an external perturber can have a variety of morphologies in inclined systems --- they may appear as one trailing arm; two trailing arms on the same side of the disk; or two arms winding in opposite directions. The disk ring outside a planetary gap may also mimic spiral arms when viewed at high inclinations. We suggest potential explanations for the features observed in HH~30, HD~141569~A, AK~Sco, HD~100546, and AB~Aur. We emphasize
that inclined views of companion-induced
features cannot be converted into face-on views using simple and commonly practiced image deprojections.

\end{abstract}



\section{Introduction}\label{sec:intro}

Gas giant planets form in gaseous disks surrounding newly born stars. As planets form in disks, their gravity perturbs the surrounding material and produces structures such as spiral density waves and gaps \citep[e.g.,][]{kley12}. These planet-induced structures are detectable in high angular resolution, direct imaging observations of disks at near-infrared (NIR) wavelengths \citep[e.g.,][]{dejuanovelar13, pinilla15twoplanets, dong15gaps, dong15spiralarms, pohl15}, which are sensitive to scattered light from the disk surface. Recently, spiral arms and gaps that are expected to result from planets have been found in disks surrounding AB Aur \citep{hashimoto11}, MWC~758 \citep{grady13, benisty15}, SAO 206462 \citep{muto12, garufi13, stolker16}, HD 100453 \citep{wagner15100453}, HD 100546 \citep{boccaletti13, avenhaus14hd100546, currie15, garufi16}, AK Sco \citep{janson16}, and TW Hya \citep{rapson15twhya}. High angular resolution, high contrast scattered light images of disks are made possible by instruments such as VLT/NACO \citep{lenzen03}, Subaru/HiCIAO \citep{tamura06}, VLT/SPHERE \citep{beuzit08}, Gemini/GPI \citep{macintosh08}, and Subaru/SCExAO \citep{jovanovic15}.

To anticipate how companion-induced disk structures may appear in NIR images, we need synthetic observations. Combining hydrodynamic simulations of disk-planet interactions with Monte Carlo radiative transfer (MCRT) calculations, \citet{dong15gaps} explored the appearance of gaps in scattered light images, and showed that the cavities in transitional disks \citep{espaillat14} can be explained as wide common gaps opened by multiple giant planets \citep{zhu11, duffell15dong}. By comparing the synthetic images of gapped disks in that study with the new GPI observations of TW Hya, \citet{rapson15twhya} concluded that the gap at 21~AU in the system may be explained by a $0.16\mj$ planet. Using a similar method, \citet{dong15spiralarms} examined the morphology of the spiral shocks excited by a giant companion in the outer disk, finding two prominent arms in a near $m=2$ rotational symmetry in scattered light images \citep[see also][]{zhu15densitywaves, fung15}, that resemble the arms discovered in MWC 758 \citep{benisty15} and SAO 206462 \citep{garufi13}. This giant-companion double-arm scenario was verified by \citet{dong16hd100453} in SPHERE images of the HD~100453 system \citep{wagner15100453}, in which an M dwarf companion is seen together with the two arms it drives.

Past models have focused on producing synthetic observations for disks viewed face-on (but see e.g., \citealt{jangcondell13}). In this work we expand the parameter space to arbitrary inclinations and position angles. We carry out three-dimensional (3D) hydro simulations using the code \texttt{PEnGUIn} \citep{fung15thesis} to calculate the density structure of spiral arms and gaps produced by a giant planet or a stellar companion on a circular orbit in the outer disk (Section~\ref{sec:simulations}). The numerical disk models are subsequently read into the \citet{whitney13} 3D Monte Carlo radiative transfer (MCRT) code to produce synthetic images at NIR wavelengths. In Section~\ref{sec:results}, we examine the morphology of the disk over a large grid of viewing geometries. In Section
\ref{sec:discussions}, after criticizing
some common reconstruction techniques for deprojecting inclined disks,
we produce synthetic observations for AB Aur and AK Sco, comparing them with their respective Subaru/HiCIAO \citep{hashimoto11} and VLT/SPHERE observations \citep{janson16}. A summary is given in Section~\ref{sec:summary}.


\section{Hydro and MCRT Simulations}\label{sec:simulations}

We carry out 3D hydrodynamics simulations to calculate disk structures induced by a companion. The resulting disk structures are fed into 3D MCRT simulations to produce synthetic images at several NIR bands. Our numerical methods are described below.

\subsection{Hydrodynamics Simulations}\label{sec:hydro}

We use the 3D Lagrangian-remap hydrodynamics code \texttt{PEnGUIn} \citep{fung15thesis} to simulate spiral arms in disks for two different models. In the first model, we have a companion-to-star mass ratio, $q$, of $0.1$. We call this the ``$0.1\msun$ model''; the companion's mass is $0.1\msun$ if the star is $1\msun$. For our
``$3 \mj$ model,'' we have $q = 0.003$. These two choices of $q$ are motivated by the three best studied double-arm systems to date: MWC~758 \citep{grady13, benisty15} and SAO~206462 \citep{muto12, garufi13, stolker16} which most likely harbor planetary mass companions with $0.001\lesssim q\lesssim0.01$ \citep{dong15spiralarms, zhu15densitywaves, fung15}, and HD 100453 \citep{wagner15100453} whose arms
are almost certainly driven by the $q=0.18$ M dwarf companion \citep{dong16hd100453}.
Both simulations are run for 100 orbits of the companion, at which point the disk structures have long settled into quasi-steady states. Except for Figure~\ref{fig:image_100mj_orbits}, all figures in this paper are produced using the models after 100 companion orbits have elapsed.

\texttt{PEnGUIn} solves the same equations as in \citet{fung15}, except in this paper all simulations are done in 3D, and with different input parameters. We provide some details here.
We use $\{r,\phi,\theta\}$ to denote the usual radial, azimuthal, and polar coordinates, and $R=r\sin\theta$ for the cylindrical radius. The Lagrangian continuity and momentum equations solved by \texttt{PEnGUIn} are:
\begin{align}
\label{eqn:cont_eqn}
\frac{D\rho}{Dt} &= -\rho\left(\nabla\cdot\mathbf{v}\right) \,,\\
\label{eqn:moment_eqn}
\frac{D\mathbf{v}}{Dt} &= -\frac{1}{\rho}\nabla p + \frac{1}{\rho}\nabla\cdot\mathbb{T}  - \nabla \Phi \,,
\end{align}
where $\rho$ is the gas density, $\mathbf{v}$ the velocity field, $p$ the gas pressure, $\mathbb{T}$ the Newtonian stress tensor, and $\Phi$ the combined gravitational potential of the star and the companion. We adopt a locally isothermal equation of state, such that $p= c^2 \rho$, where $c$ is the sound speed of the gas. $\mathbb{T}$ is proportional to the kinematic viscosity $\nu$, which we parameterize using the $\alpha$-prescription by \citet{shakura73}, such that $\nu=\alpha c^2/\Omega$, where $\Omega$ is the orbital frequency of the disk. In all our simulations, we choose $\alpha=0.01$. This relatively high viscosity is chosen to isolate the spirals excited by the companion from additional disk features caused by hydrodynamical instabilities, such as the Rossby wave instability \citep{lovelace99} triggered by the formation of a sharp gap edge. A higher viscosity also shortens the system's viscous timescale, allowing the simulation to converge faster with time.

The simulations are performed in the frame of the central star, so for a companion on a fixed, circular orbit, $\Phi$ is:
\begin{equation}\label{eqn:potential}
\Phi = -GM\left[\frac{1-q}{r}-\frac{q}{\sqrt{r^2 + r_{\rm s}^2 + \rp^2 - 2R\rp\cos{\phi'}}} - \frac{qR\cos{\phi'}}{\rp^2}\right] \,,
\end{equation}
where $G$ is the gravitational constant, $M$ the total mass of the star and the companion, $\rp$ the semi-major axis of the companion's orbit, $r_{\rm s}$ the softening length of the companion's potential, and $\phi' = \phi-\phi_{\rm p}$ denotes the azimuthal separation from the companion. For convenience, we also denote the Keplerian velocity and frequency as $v_{\rm k}$ and $\Omega_{\rm k}$. The third term in the bracket is the indirect potential due to the acceleration of the frame. $r_{\rm s}$ is chosen to be $10\%$ of the companion's Hill radius, so for the $0.1 \msun$ model, we have $r_{\rm s} = 0.032 \rp$, and for the $3 \mj$ model, $r_{\rm s} = 0.01 \rp$. We note that the companion is introduced into the simulation gradually --- $q$ increases from zero to the desired value over 5 orbits.

Because we adopt different disk sound speed profiles for the two models (see the following section), we adjust our simulation domains and resolutions accordingly. In the $0.1 \msun$ model, our simulation box spans $\rin=0.1 \rp$ to $\rout=2 \rp$ in radius, the full $2\pi$ in azimuth, and $0^\circ$ to $39^\circ$ in the polar angle measured from the midplane (symmetry is enforced across the midplane). The simulation grid contains $216(r)\times432(\phi)\times72(\theta)$ cells, with the radial cells spaced logarithmically, while the spacing is uniform for the other two dimensions. For the $3 \mj$ model, our simulation box spans $0.2 \rp$ to $2 \rp$ in radius, the full $2\pi$ in azimuth, and $0^\circ$ to $31^\circ$ in the polar direction. The resolution is higher in this model with $288(r)\times576(\phi)\times96(\theta)$ cells, because of the choice of a lower sound speed.

\subsubsection{Initial and boundary conditions}\label{sec:initial}
The initial disk profile assumes a $\Sigma\propto1/r$ surface density profile. Additionally, we impose an initial axisymmetric gap that mimics the gap we expect the companion to open. The inclusion of this initial gap allows for a significantly faster time-convergence, and avoids some numerical artifacts that would be caused by the sudden introduction of a companion into an unperturbed disk. The initial density profile reads:
\begin{equation}\label{eqn:initial_rho}
\rho = \rho_0 \left(\frac{R}{\rp}\right)^{-\frac{5}{2}+\beta} \exp\left(\frac{GM}{c^2} \left[\frac{1}{r}-\frac{1}{R}\right]\right) \left(\frac{1}{2} - \frac{1}{\pi}\tan^{-1}\left[\frac{R_{\rm gap}-|R-\rp|}{\Delta R_{\rm gap}}\right]\right)\,,
\end{equation}
where $\rho_0 = 1$,\footnote{Since we do not consider the self-gravity of the disk, the normalization $\rho_0$ has no impact on our results.} $\beta$ describes the disk sound speed profile, $R_{\rm gap}$ is the distance from $\rp$ to the gap edge, and $\Delta R_{\rm gap}$ describes the sharpness of the gap edge. We choose $R_{\rm gap}$ to be approximately two times the companion's Hill radius \citep{fung14}, which makes $R_{\rm gap} = 0.6$ for the $0.1 \msun$ model, and $R_{\rm gap} = 0.2$ for $3 \mj$. We choose $\Delta R_{\rm gap} = 0.08$ in both models, which is sufficiently smooth to not trigger any hydrodynamical instabilities. The sound speed profile reads:
\begin{equation}\label{eqn:initial_c}
c = c_0 \left(\frac{R}{\rp}\right)^{-\beta} \,,
\end{equation}
where $c_0$ is the sound speed at the companion's radial position. For the $0.1 \msun$ model, we choose $c_0 = 0.15 v_{\rm k,p}$, where $v_{\rm k,p}$ is the Keplerian speed at $R=\rp$, and $\beta = 1/2$. This constitutes a constant disk aspect ratio of $c/v_{\rm k} = 0.15$. For the $3 \mj$ model, we choose $c_0 = 0.1 v_{\rm k,p}$ and $\beta = 1/4$, which makes a flaring disk with $c/v_{\rm k} \propto R^{1/4}$. The somewhat larger scale height for the $0.1\msun$ model is motivated by the recent fits to HD~100453 \citep{dong16hd100453}, and also provides a contrast to the $3\mj$ model.

The initial velocity field assumes hydrostatic equilibrium taking into account gas pressure, but neglecting viscosity. As a result, the initial disk has zero radial and polar velocities, and an orbital frequency of:
\begin{equation}\label{eqn:omega}
\Omega = \sqrt{\Omega_{\rm k}^2 + \frac{1}{r}\frac{\partial p}{\partial r}} \,,
\end{equation}

Both the inner and outer radial boundaries are fixed at their initial values. For the polar boundaries, we enforce symmetry at the midplane since we only simulate the upper half of the disk, and use a reflecting condition at the top to prevent mass from entering the simulation domain. Additionally, a wave killing zone is imposed next to the top boundary to remove reflections. The prescription for this wave killing zone is:
\begin{equation}\label{eqn:kill}
\frac{\partial{X}}{\partial t} = -X \Omega_{\rm k} \left(1-\frac{\theta_{\rm top}-\theta}{\Delta\theta}\right)^2 \, ,
\end{equation}
where $X$ represents the fluid variable $\rho$, $p$, or $\mathbf{v}$, $\theta_{\rm top}$ is the polar angle at the top boundary, and $\Delta\theta$ is the width of the killing zone, which is $5^\circ$ for the $0.1 \msun$ model and $3^\circ$ for $3 \mj$. Technically, this killing zone results in a continuous loss of disk mass, but since the top boundary is located about 5 scale heights above the midplane, the loss is negligible over the timescale of our simulations.

\subsection{Monte Carlo Radiative Transfer Simulations}\label{sec:mcrt}

The density distributions obtained in our hydro simulations are post-processed via 3D MCRT calculations using the \citet{whitney13} code to produce synthetic NIR images. This process largely follows the procedures described in \citet{dong15spiralarms} and \citet{dong16hd100453}. The central source is assumed to be a protostar with a temperature of 5200~K and a radius of $2.4R_\odot$ (MCRT simulations have only one illumination source, i.e., the central star). The temperature in each grid cell is calculated using the radiative equilibrium algorithm described in \citet{lucy99}. The systems are assumed to be at 140 pc from the Earth, the distance to the Taurus star forming region. The companion is assumed to be at $\rp=100$ AU in the $3\mj$ case and 200 AU in the $0.1\msun$ case, chosen so that the spiral arms in both cases are most prominent around 50 to 100 AU, as is suitable for NIR imaging observations with currently achievable inner working angles and angular resolutions for systems at 140 pc. These separations are also in the same range as the distances to the (predicted) companions in SAO 206462, MWC 758, and HD 100453. The model grid is a 3D spherical grid in $\{r,\phi,\theta\}$. In the $\phi$ and $\theta$ directions the grid is identical to the one in the corresponding hydro simulation. In the $r$ direction, it extends from the dust sublimation radius $\rsub$, where the dust temperature reaches the sublimation temperature of 1600 K, to $\rout$. From $\rsub$ to $\rin$ the grid has 50 cells in $r$ with logarithmic spacing, while from $\rin$ to $\rout$ the grid is identical to the corresponding hydro simulation. 

Since hydro simulations produce disk density structures in gas, while NIR images are determined by the distribution of dust in disks, we need to convert gas density structures into dust density structures. The dust grains are assumed to be interstellar medium (ISM) grains \citep{kim94} made of silicate, graphite, and amorphous carbon. Their size distribution obeys a smooth power law up to a grain radius of 0.25~$\micron$ followed by an exponential cut off at larger sizes. The optical properties of the dust grains can be found in \citet[][Figure 2]{dong12cavity}. From $\rin$ to $\rout$, the dust volume density, $\rhodust$, is assumed to be linearly proportional to the gas volume density, $\rhogas$. This is appropriate as these sub-$\micron$-sized particles are dynamically well coupled to the gas. Between $\rsub$ and $\rin$, $\rhodust$ at $(r,\phi,\theta)$ is assumed to be the same as $\rhodust$ at $(\rin,\phi,\theta)$. We note that this inner disk is introduced in the MCRT simulations simply to fill in the central cavity of the hydro simulations. Removing this inner disk leads to a $\lesssim35\%$ increase in the surface brightness of the resultant image, but does
not alter our reported outer disk morphologies. The total ISM dust disk mass within 100 AU is normalized to $10^{-5}\msun$ in both models. This corresponds to, for example, a total gas disk mass of 0.01~$\msun$, a 100:1 gas-to-dust mass ratio, a $10\%$ dust mass fraction in the small ISM grains, and a remaining 90\% mass fraction in the large grains that are presumed to have settled to the disk midplane and do not affect NIR scattering. Tests show that our results are insensitive to the total dust mass within one order of magnitude around this fiducial value.

Our MCRT simulations output all 4 Stokes components, $(I,Q,U,V)$, from which total intensity (TI=$I$) and polarized intensity (PI=$\sqrt{Q^2+U^2}$) images\footnote{In this work, the physical quantity recorded in all model images is the specific intensity in unit of [mJy~arcsec$^{-2}$], or [10$^{-26}$ ergs~s$^{-1}$~cm$^{-2}$~Hz$^{-1}$~arcsec$^{-2}$].} are produced at $Y$, $J$, $H$, and $K$-band using 1 billion photon packets over a range of viewing angles. Polarization fraction (PF) is defined as PF$=$PI/TI. In addition, the following polar-coordinate Stokes parameters are defined \citep{schmid06}:
\begin{equation}
Q_r=+Q\cos{2\phi}+U\sin{2\phi},
\label{eq:qr}
\end{equation}
\begin{equation}
U_r=-Q\sin{2\phi}+U\cos{2\phi},
\label{eq:ur}
\end{equation}
where $\phi$ is the position angle of a point (North to East) in a disk image. By construction, $Q_r$ contains the polarization components in the tangential (azimuthal) and radial directions, while $U_r$ contains the components at $\pm45^\circ$ with respect to that. If the disk scattering surface is perpendicular to the line of sight (this is approximately true for geometrically thin disks viewed face-on) and observed photons are mostly produced by single-scattering events, only the tangential polarization component exists, and $Q_r$ is equivalent to PI (except unlike PI, $Q_r$ can be both positive and negative). In such cases, $U_r$ is expected to contain little signal and is often used to estimate the noise level \citep{quanz13gap, garufi13, canovas13, pinilla15j1604, rapson15twhya}. For inclined disks (for which the scattering angle is often not $90^\circ$) and/or images built up of multiply scattered photons (as is the case for disks surrounded by an envelope), polarization may not be purely tangential, and $U_r$ can contain non-trivial signal as well \citep{garufi16}. This issue has been recently investigated by \citet{canovas15qrur}, and will be discussed in Section~\ref{sec:results_arms}.

The orientation of the disk inclination is indicated by a pair of parameters (PA,$i$), where PA is the position angle of the disk major axis (North-to-East; PA$=0^\circ$ when the west side of the disk is the nearside) and $i$ is the inclination ($0^\circ$ is face-on and $90^\circ$ is edge-on). The images emerging from the MCRT pipeline are convolved by a Gaussian point spread function (PSF) with a full width half maximum (FWHM) of $0.025\arcsec$ at $Y$-band, $0.03\arcsec$ at $J$-band, $0.04\arcsec$ at $H$-band, and $0.05\arcsec$ at $K$-band. These convolving PSFs are chosen to achieve the diffraction limited angular resolution at each wavelength for an 8-meter telescope (e.g. Subaru, VLT, or Gemini).


\section{Modeling Results}\label{sec:results}

In this section, we present the morphology of the spiral arms in both the $3\mj$ and $0.1\msun$ models, and the morphology of the gap in the $3\mj$ model (in the $0.1\msun$ model the disk is effectively truncated to a circumprimary disk), over the full range of viewing angles. Our hydro simulations settle into quasi-steady states after about 10 orbits, and the morphology of the arms remains practically the same afterwards, as shown in Figure~\ref{fig:image_100mj_orbits}.

\begin{figure}
\begin{center}
\includegraphics[trim=0 0 0 0, clip,width=0.5\textwidth,angle=0]{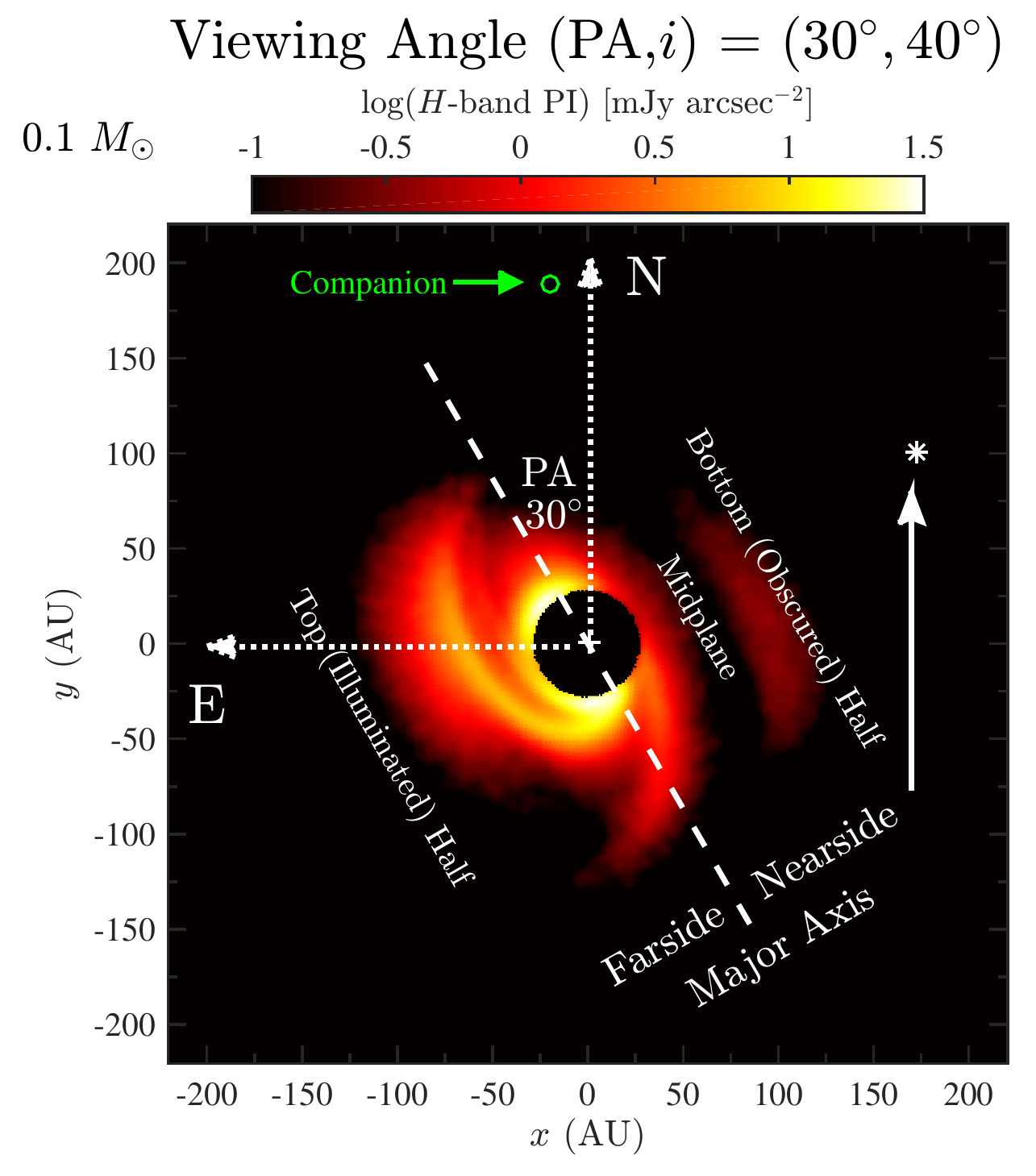}
\end{center}
\figcaption{Definitions of geometrical terms and figure notations used throughout this paper. The figure is the convolved $H$-band PI image for the $0.1\msun$ model viewed at (PA,$i$)=($30^\circ,40^\circ$). In all images like this one in this paper, North is up and East is left. The projected location of the companion is indicated by the open green circle. 
The major axis of the disk, which always
passes through the star, is marked by the dashed line. 
The nearside/farside of the disk are divided by the major axis, while the top (illuminated) half and the bottom (obscured) half of the disk are separated by the dark lane at the disk midplane.
The nearside of the disk is marked by the white asterisk on the minor axis. The position angle PA is defined as the North-to-East angle of the major axis. In this paper PA lies in the range of [$-180^\circ,180^\circ$], and is positive when the nearside of the disk lies to the North. 
\label{fig:image_geometry}}
\end{figure}

The definitions of various geometrical terms used in this paper, including the disk's
PA; its ``top'' (illuminated) and ``bottom'' (obscured) halves; and its ``nearside'' and ``farside,'' are illustrated in Figure~\ref{fig:image_geometry}.
We use a ``red-hot'' color scheme for the scattered light images and a greyscale  scheme for other plots. Unless stated otherwise, all MCRT images are convolved polarized intensity images at $H$-band (angular resolution $0.04\arcsec$;
we also examined the arm morphologies at $Y$, $J$, $H$, and $K$-bands, and found them indistinguishable);
the central $0.2\arcsec$ in radius is masked out to mimic a coronagraph; the star is marked by a white $+$; the nearside of the disk is indicated by a white asterisk on the minor axis; and the projected location of the companion is indicated by an open green circle (the absence of the green circle implies that the companion is outside the field of view). We note that the noise level (i.e. the detection limit) in current NIR PI imaging observations is $\sim$0.1 mJy arcsec$^{-2}$ or lower \citep[e.g.,][]{hashimoto12, mayama12, kusakabe12, grady13, follette13}.\footnote{0.1 mJy arcsec$^{-2}$ is the detection limit for AO188+HiCIAO onboard Subaru. The detection limits for the newer generation of instruments, such as Gemini/GPI and VLT/SPHERE, are expected
to be better.} Our color scheme is chosen
to bring out all features above
this noise floor.

We also experimented with varying the total disk dust mass $M_{\rm dust}$ by 1 order of magnitude around our fiducial value, and did not find any noticeable change in our results. Since different frequency bands and $M_{\rm dust}$ values probe scattering surfaces of different altitudes, this insensitivity suggests that the positions of the arms are largely independent of altitude.

\subsection{Face-on Morphologies}\label{sec:results_faceon}

Figure~\ref{fig:basic} shows the surface density maps and face-on images (both full resolution and convolved) for both models. The companion in both cases excites two spiral shocks inside its orbit, which manifest themselves as arms in scattered light images, as seen in previous works \citep{dong15spiralarms, zhu15densitywaves, fung15, dong16hd100453}. We will label as primary (``P'') the arm pointing towards the companion; the other arm is secondary (``S''). The two arms in the stellar mass companion case are in a near $m=2$ rotational symmetry. In the
planetary companion case, the arms are $\sim$$130^\circ$ apart, consistent with the relation found by \citet{fung15}. Also, because of the smaller scale height in the $3\mj$ model, the spiral arms in that model are less open than in the $0.1\msun$ model. The loci of the arms are marked in Figure~\ref{fig:image_locus}.

The $3\mj$ planet opens a gap from 75 to 130 AU. Both the inner disk, which harbors the prominent inner arms, and the outer disk ring are visible in scattered light. The planet also excites two arms in the outer disk ring; however they are faint because of their large distance from the star, and they are difficult to distinguish 
because they are tightly wound \citep{dong15spiralarms}. In this paper we focus on the inner arms as they are most likely to be observable. 

In the $0.1\msun$ model, the powerful tidal force of the companion truncates the disk to about half its orbital radius (100~AU), leaving behind two separate components --- a circumprimary disk and a circumsecondary disk. We note that the circumsecondary disk also has a pair of mini-arms excited by the primary star; however, since the companion is not an illumination source, this mini-disk is not visible in scattered light (it is shadowed from the primary star by the circumprimary disk).

\subsection{The Inner Arms at Various Viewing Angles}\label{sec:results_arms} 

Figures~\ref{fig:image_100mj_geometry},
\ref{fig:image_100mj_geometry_big},
and \ref{fig:image_3mj_geometry} show synthetic $H$-band images for the two models at various viewing angles (Figure~\ref{fig:image_100mj_geometry_big} is a zoomed-out version of Figure~\ref{fig:image_100mj_geometry} that shows the location of the companion). For the $3\mj$ model, we view the disk at 7 inclination angles from $20^\circ$ (at $i<20^\circ$ disk images are very similar to face-on; see Figure~\ref{fig:basic}) to $80^\circ$ (top to bottom); at each inclination the disk is oriented at 12 PAs covering the entire $2\pi$. In total, $7\times12=84$ images of different viewing angles are shown. For the $0.1\msun$ model, because the two arms are nearly rotation-symmetric, at each inclination we show the disk at 6 PAs covering $\pi$ in the azimuthal direction (the other $\pi$ is redundant), and thus, in total, $7\times6=42$ images of different viewing angles are shown.

The distortion of the arms introduced by a non-face-on viewing angle can be dramatic. In general, the portion of the arms on the nearside of the disk gradually moves towards the major axis as the inclination increases. This can be seen at all position angles, and is best illustrated when an entire arm is on the nearside (e.g., from top to bottom in the PA$=0^\circ$ column in Figure~\ref{fig:image_100mj_geometry}, and
from top to bottom in the PA$=90^\circ$ and $-180^\circ$ columns in Figure~\ref{fig:image_3mj_geometry}). This behavior is expected when viewing a bowl-shaped surface at an angle. The distortions as seen on the farside of the disk are less dramatic, until the disk reaches very high inclination ($\gtrsim70^\circ$) and the entire top half becomes a bowl viewed edge-on. Because of the distortions generated by finite inclinations, what appear as two trailing arms on two sides of the disk in face-on images may appear to be (1) just one arm; (2) two trailing arms on one side of the disk; or (3) two arms on one side winding in opposite directions. These possibilities are summarized in Figure~\ref{fig:image_100mj_pianalysis}, which shows the $0.1\msun$ model at four viewing angles.

At inclinations lower than $\sim40^\circ$,
the disk's bottom half is too faint to
detect (its emission is less than three
times the noise level in current
observations). 
As the disk becomes more inclined, the
nearside of its bottom half,
separated from the top half by the dark
lane at the disk midplane,
brightens.\footnote{The specific inclination threshold for the bottom half to emerge depends on the total mass of the small grains: for a smaller total dust mass, the disk is less optically thick, and thus the dark lane at the midplane is thinner, rendering the bottom half more easily seen.} In the $0.1\msun$ model, because there is no outer disk to block the spiral arms, one or both arms can be seen on the disk's bottom half depending on the position angle. This is best illustrated by comparing the first and fourth columns of Figure~\ref{fig:image_100mj_geometry_big}. In the $3\mj$ model, only the outer disk ring can be seen on the bottom half, as the
outer disk blocks the inner disk. Therefore the bottom half remains symmetric about the minor axis, regardless of position angle.

$Q_r$, $U_r$, and polarization fractions for various viewing angles are shown in Figure~\ref{fig:image_100mj_pianalysis}, along with TI and PI images. PI and $Q_r$ are nearly identical in all cases, and
aside from a difference in normalization,
both resemble the TI images. While the face-on $U_r$ contains little signal and represents noise in our MCRT simulations, at high inclinations there is some signal in $U_r$ (particularly in the bright inner disk close to the model coronagraph edge), which shows patterns symmetric about the minor axis. These findings are consistent with those by \citet{canovas15qrur} for disks
viewed at modest-to-high inclination.

\subsection{The Outer Disk and the Gap in the $3\mj$ Case at Various Inclinations}\label{sec:results_gap}

The $0.1\msun$ disk is truncated by the companion and thus has no outer disk, while in the $3\mj$ case the planet opens a gap that separates the well-defined inner and outer disks. Figure~\ref{fig:image_3mj_geometry_gap} shows PI, TI, and PF images of the entire 200-AU-wide disk for the $3\mj$ model at 
PA$=90^\circ$ for
various inclinations (the PF images are the same as in the first column of Figure~\ref{fig:image_3mj_geometry}). The outer disk ring on the top half is visible in both PI and TI at all inclinations. In TI images, scattered light is brightest on the nearside of the ring, and decreases towards the farside because of preferential forward scattering by ISM dust. In PI images, both the nearside and farside of the top half are fainter than in the TI images because of the inefficiency of polarized scattering at both small and large scattering angles (note the two breaks on the ring along the minor axis at $i=50^\circ$). The two ``wings'' along the major axis on the top half of the disk are prominent as the scattering angle is close to $90^\circ$. At high inclinations, the two wings may look like two spiral arms winding in opposite directions (as further discussed in Section~\ref{sec:discussions_aksco}). This can be further demonstrated in the right column: the PF in the two wings (indicated by the red arrows) at $i\gtrsim60^\circ$ can reach $\sim$0.5, while the farside of the top half and the nearside of the bottom half have PF close to 0. The gap is traceable in both TI and PI images at $i\lesssim60^\circ$ (at larger inclinations only the TI images show the gap), but leaves little imprint in the PF maps. Lastly, the dark lane of the disk midplane is prominent at $i\gtrsim60^\circ$ in both TI and PI, but is practically 0 in PF.

Figure~\ref{fig:image_3mj_edgeon} shows the TI and PI images at 3 nearly edge-on viewing angles ($i=80^\circ$) in a linear stretch to emphasize the structures in the outer disk ring. In all cases, clumps can be identified on the ``wings'' in both TI and PI. They are spiral arms ``collapsed'' along the line of sight. These structures will be discussed in Section~\ref{sec:discussions_aksco}.


\section{Discussion}\label{sec:discussions}

\subsection{The Effects of Deprojecting and $r^2$-Scaling Inclined Disk Images}\label{sec:discussions_deproj}

In resolved observations of disks, the original images are sometimes deprojected based on an estimated inclination of the system in an attempt to recover the face-on morphology of the disk, and also scaled by $r^2$ to enhance the visibility of features at large distance (sometimes the two are combined as in, e.g., \citealt{thalmann15}). Clearly, deprojection assumes the disk surface is thin and flat (i.e., not curved), and $r^2$-scaling assumes a face-on viewing geometry. Neither of these conditions is typically satisfied. To assess the degree to which conventional
deprojection and scaling procedures introduce unphysical artifacts, we show in Figures~\ref{fig:image_3mj_deproj} and \ref{fig:image_100mj_deproj} the results of deprojections and scalings. $r^2$-scaling significantly enhances the visibility of features at large projected distances, as expected. In particular, the originally faint bottom half of the disk becomes prominent, as does the ring in the outer disk in the $3\mj$ model. However we note that the compensation at large radius is not entirely physical. For example, different parts of the outer disk ring in the $3\mj$ model are not compensated by the same factor (which they should be because the ring is a circle centered on the star) because of their different projected distance $r$ from the center. As a result, the ring is disproportionally enhanced along the major axis and less so along the minor axis.

Likewise, deprojecting an inclined image can sometimes lead to serious misinterpretation of the true morphology of the system. In the $0.1\msun$ model, even at a modest $30^\circ$ inclination (the second and third rows in Figure~\ref{fig:image_100mj_deproj}), deprojection cannot restore the morphology of the two arms at a face-on viewing angle; at $50^\circ$ inclination (the last two rows), deprojection severely distorts the arms --- it cannot even recover the total number of arms. In the $3\mj$ model, deprojection also fails to restore the face-on arm morphologies, although the distortion is less severe than for the $0.1\msun$ model because the disk is flatter in the $3\mj$ model ($h/r=0.1$) than in the $0.1\msun$ model ($h/r=0.15$), and because the height variations of the arms are smaller for the $3\mj$ model. Deprojection also cannot restore the outer disk ring and the gap back to concentric circular structures --- they appear to be off-centered ellipses broken along the minor axis (the eccentricity of the ellipses is somewhat reduced compared with the original inclined images). Lastly, it is obvious that the dark lane along the disk midplane cannot be removed by deprojecting the images.

It follows from these tests that the method developed by \citet{fung15} to infer the mass of a planet based on the azimuthal separation between the two arms it excites can only be applied to nearly face-on systems ($i\lesssim20^\circ$). 

\subsection{Connections with Observations}\label{sec:discussions_obs}

In this section, we connect our models to observed disk systems.

\subsubsection{Inhomogeneities in Nearly Edge-on Disks}\label{sec:discussions_aumic}

As shown in the bottom rows of Figures~\ref{fig:image_100mj_geometry},
\ref{fig:image_100mj_geometry_big},
and \ref{fig:image_3mj_geometry}, nearly edge-on disks ($i\sim80^\circ$) with spiral arms may appear to be asymmetric about their minor axis at certain PAs, as the axisymmetry of the disk is broken by the non-axisymmetric spiral arms excited by companions. In the case of a stellar mass companion, the spiral arms are big and wide, and not blocked by an outer disk ring when viewed edge-on (since the disk is severely truncated); the arms introduce global asymmetries in disks. This kind of asymmetry has been found in directly imaged nearly edge-on disks, such as the HH~30 disk inclined by $84^\circ$ \citep{stapelfeldt99}.\footnote{http://hubblesite.org/newscenter/archive/releases/2000/32/image/c/} In the case of planet-induced spiral arms (which are thinner), when viewed at certain PAs they may ``collapse'' along the line of sight to produce intensity inhomogeneities in the form of clumps in edge-on disks (Figure~\ref{fig:image_3mj_edgeon}). These clumps can be seen in TI and especially PI images. In direct imaging observations, they may mimic and be confused with point source detections. In recent GPI and SPHERE observations of the edge-on debris disk AU Mic, several clumps of unknown origin have been identified at tens of AU on one side of the disk \citep{wang15, boccaletti15}. The morphology of these clumps are reminiscent of the ones in Figure~\ref{fig:image_3mj_edgeon}. Though our results cannot be directly applied to the AU Mic observations as gas-poor debris disks differ in their dynamics from gas-rich protoplanetary disks, Figure~\ref{fig:image_3mj_edgeon} hints at a potential solution to the mystery, and warrants more specific investigation in the future.

\subsubsection{Partial Gaps in Disks at Low-to-Modest Inclinations}\label{sec:discussions_abaur}

As shown in Figure~\ref{fig:image_3mj_geometry_gap} a planet-induced gap may be detectable in both TI and PI images at inclinations between $\sim$$20^\circ$ and $\sim$$50^\circ$. In Figure~\ref{fig:image_abaur} we compare our $3\mj$ model at $i=20^\circ$ with the Subaru/HiCIAO polarized observation of AB Aur \citep{hashimoto11}. The full resolution model image is rescaled to match the size of the object (the $3\mj$ planet is now at 67 AU), and convolved by an appropriate PSF to achieve the same angular resolution as the Subaru observation. The model disk has PA$=-135^\circ$, $i=20^\circ$; rotates around the star in the counterclockwise direction; and has a nearside to the southeast; all these parameters are consistent with observations \citep{fukagawa04, hashimoto11, tang12}.\footnote{PA=$-135^\circ$ in this paper is equivalent to PA=$45^\circ$ in \citet{hashimoto11}; there is a difference of $\pi$ between our definitions.} The qualitative morphologies of the inner disk, the gap, and the outer disk ring resemble features seen in AB Aur; in particular, the gap at $r\sim85$ AU appears to be deeper on the farside and shallower on the nearside because of the variation of dust scattering efficiency with scattering angle. We note that although the comparison between our model and the observations is only qualitative, the model is ``unique'' in the sense that the viewing geometry has been fixed by the observations. There are even hints of fine spiral arms in the inner disk in the Subaru image, at the same locations as in the model images. We note that in the model image the morphologies of the outer ring and the gap are more-or-less independent of the position angle of the planet as they are intrinsically axisymmetric, while the details of the inner disk structures are obviously sensitive to planet location.
 
\subsubsection{The Variety of Spiral Arms in Disks}\label{sec:discussions_aksco}

As shown in Figures~\ref{fig:image_100mj_geometry}--\ref{fig:image_100mj_pianalysis}, depending on the orientation of the disk, the trailing-double-arm pattern for a face-on viewing geometry may appear at other viewing angles to be one trailing arm, two trailing arms to one side, or two arms winding in opposite directions. Such features have been reported in real disks, as in the single trailing arm at $\sim$130~AU in the inner disk of HD 141569 A \citep[viewed at $\sim$$55^\circ$ inclination;][]{konishi16}.\footnote{Although HD 141569 A is considered by some to be a debris disk, recent PdBI observations by \citet{pericaud14} found a large amount of CO gas in this $\sim$5-Myr-old system \citep{weinberger00}.} Companion-induced spiral arms viewed at a large inclination may explain such an observation.

Intriguingly, ``spiral arms'' can be mimicked by an outer disk ring separated from the inner disk by a planetary gap when viewed at non-zero inclinations, as illustrated in Figure~\ref{fig:image_3mj_geometry_gap}. In this case, the outer disk manifests itself as two ``pseudo-arms'' along the major axis, approximately symmetric about the minor axis, winding in opposite directions. The two arms found in recent SPHERE observations of AK Sco \citep{janson16} may provide a real-life example. In Figure~\ref{fig:image_aksco}, we compare a model image (rescaled to match the size of AK Sco) with the SPHERE observations. The model image adopts the geometry of AK Sco: the position angle of the disk\footnote{PA=$-130^\circ$ in this paper is equivalent to PA=$50^\circ$ in \citet{czekala15}; there is a difference of $\pi$ between our definitions.} is $-130^\circ$ and the disk is inclined by $70^\circ$ \citep{czekala15}. The two pseudo-arms in the model image qualitatively resemble the arms in AK Sco. As was the case above for AB~Aur, the model has no free parameter regarding viewing geometry. Moreover, the only remaining free parameter, the location of the planet, does not much affect the arm morphologies. Experiments show that the opening angle of the pseudo-arms is determined by the inclination (Figure~\ref{fig:image_3mj_geometry_gap}) and the disk scale height (both the absolute value and the extent of flaring). For a given distance to the object, the length of the arms is set by the size of the outer disk ring; for a fixed ring size, the width of the arms is set by the gap size determined by $\mplanet$, $h/r$, and disk viscosity \citep{fung14, duffell15gap}, as a wider gap narrows the outer disk ring. The $3\mj$ planet in our model of AK Sco is located at 25 AU from the star, while its azimuthal location cannot be determined as the outer disk ring is approximately axisymmetric.

Our model makes a unique prediction for such pseudo-arms. As shown in Figure~\ref{fig:image_3mj_geometry_gap}, they have high polarization fractions because they scatter light at $90^\circ$. Given that the SPHERE observations in \citet{janson16} are in TI, future PI observations of AK Sco can test our interpretation. Also, deeper exposures can attempt to reveal the faint nearside of the bottom half of the disk, parallel to the two arms and separated by the dark midplane.

The two major arms identified in the inner disk of HD 100546 at $r\sim40$~AU in recent GPI \citep{currie15} and SPHERE imaging \citep[][see the ``wing structure''  in their fig. 3]{garufi16} might also be pseudo-arms. The geometry of the disk supports this interpretation: the disk is suitably inclined by $\sim$$45^\circ$, and both arms are on the farside of the major axis. The above tests outlined for AK Sco also apply to HD 100546.


\section{Summary}\label{sec:summary}

\begin{figure}
\begin{center}
\includegraphics[trim=0 0 0 0, clip,width=\textwidth,angle=0]{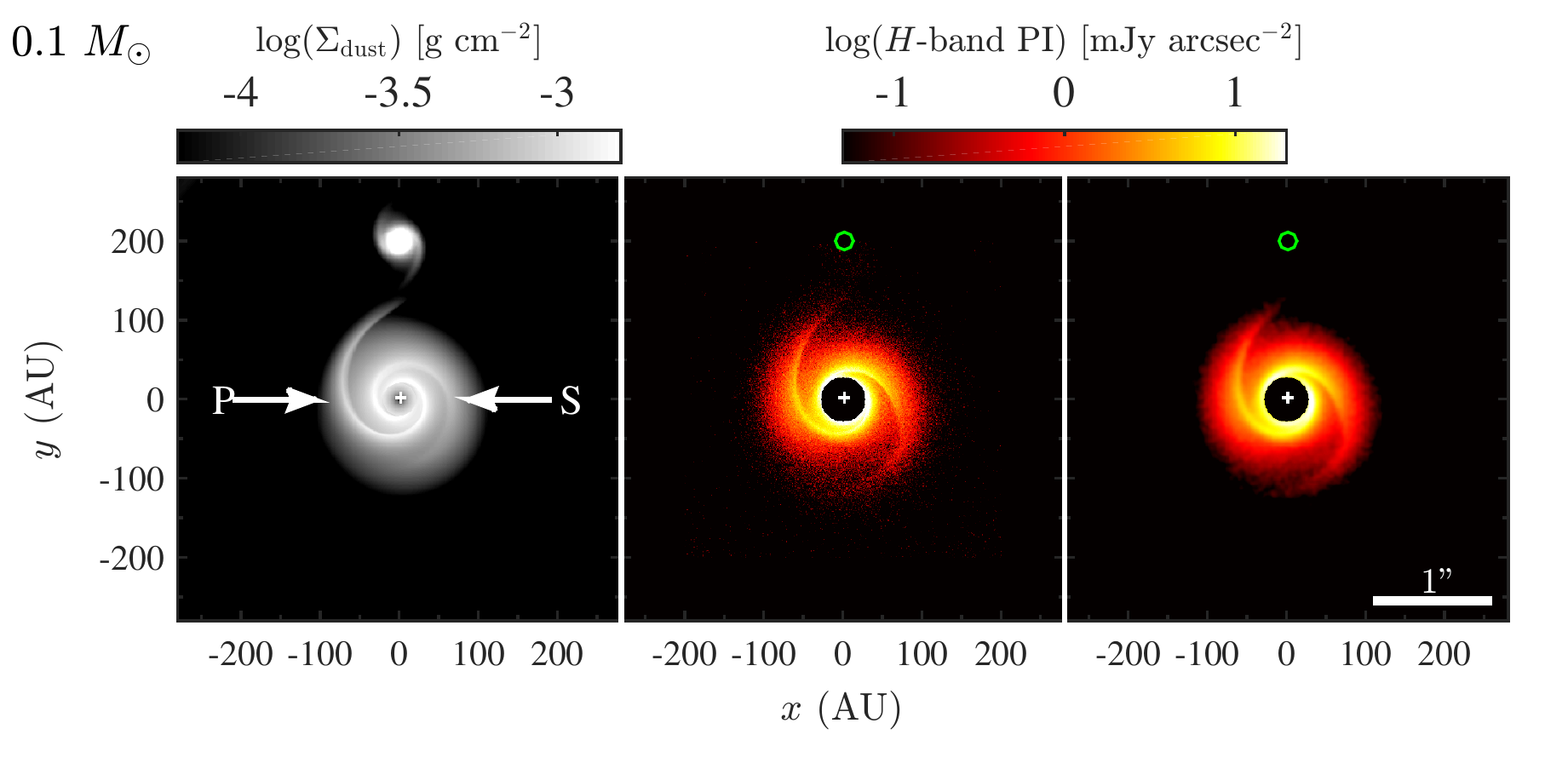}
\includegraphics[trim=0 0 0 0, clip,width=\textwidth,angle=0]{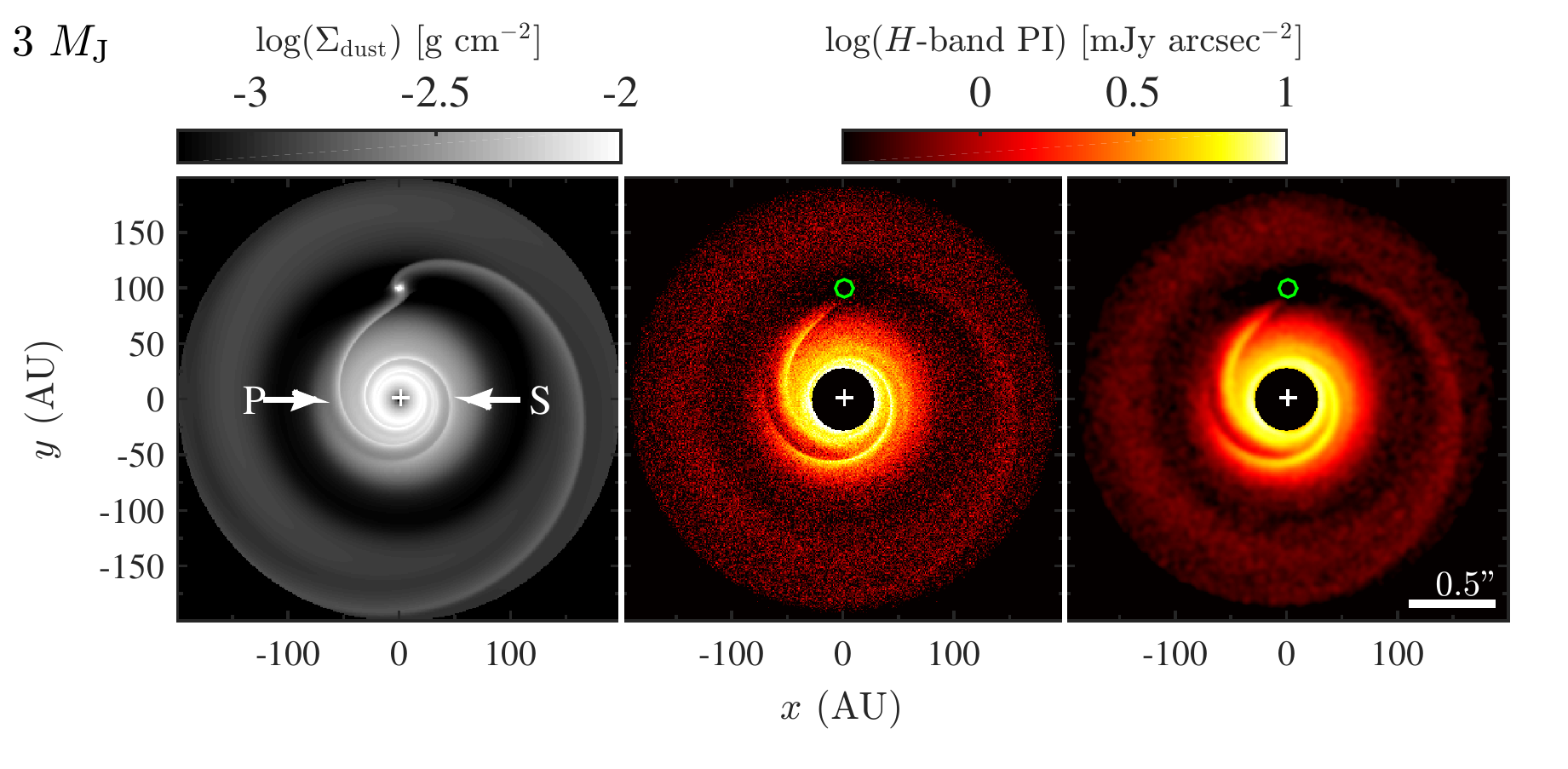}
\end{center}
\figcaption{Surface density of the small ISM-like grains (left), full resolution polarized intensity MCRT image (middle), and convolved image (right; angular resolution $0.04\arcsec$) in log scale at $H$-band for our two models. ``P'' and ``S'' in the surface density panel indicate the primary and secondary arms. The central $0.2\arcsec$ in radius in $H$-band images is masked out to mimic a coronagraph. The location of the companion (i.e., the circumsecondary disk) is evident in the surface density map, and is indicated by the open green circle in the $H$-band images. The sensitivity level in current NIR PI imaging observations is below 0.1~mJy arcsec$^{-2}$. The disk and the companion rotate around the star counterclockwise. Both the inner and outer disk (and the arms within) are visible in the $3\mj$ case, separated by a gap; only the inner disk (and the arms within) can be seen in the $0.1\msun$ case.
\label{fig:basic}}
\end{figure}

\begin{figure}
\begin{center}
\includegraphics[trim=0 0 0 0, clip,width=\textwidth,angle=0]{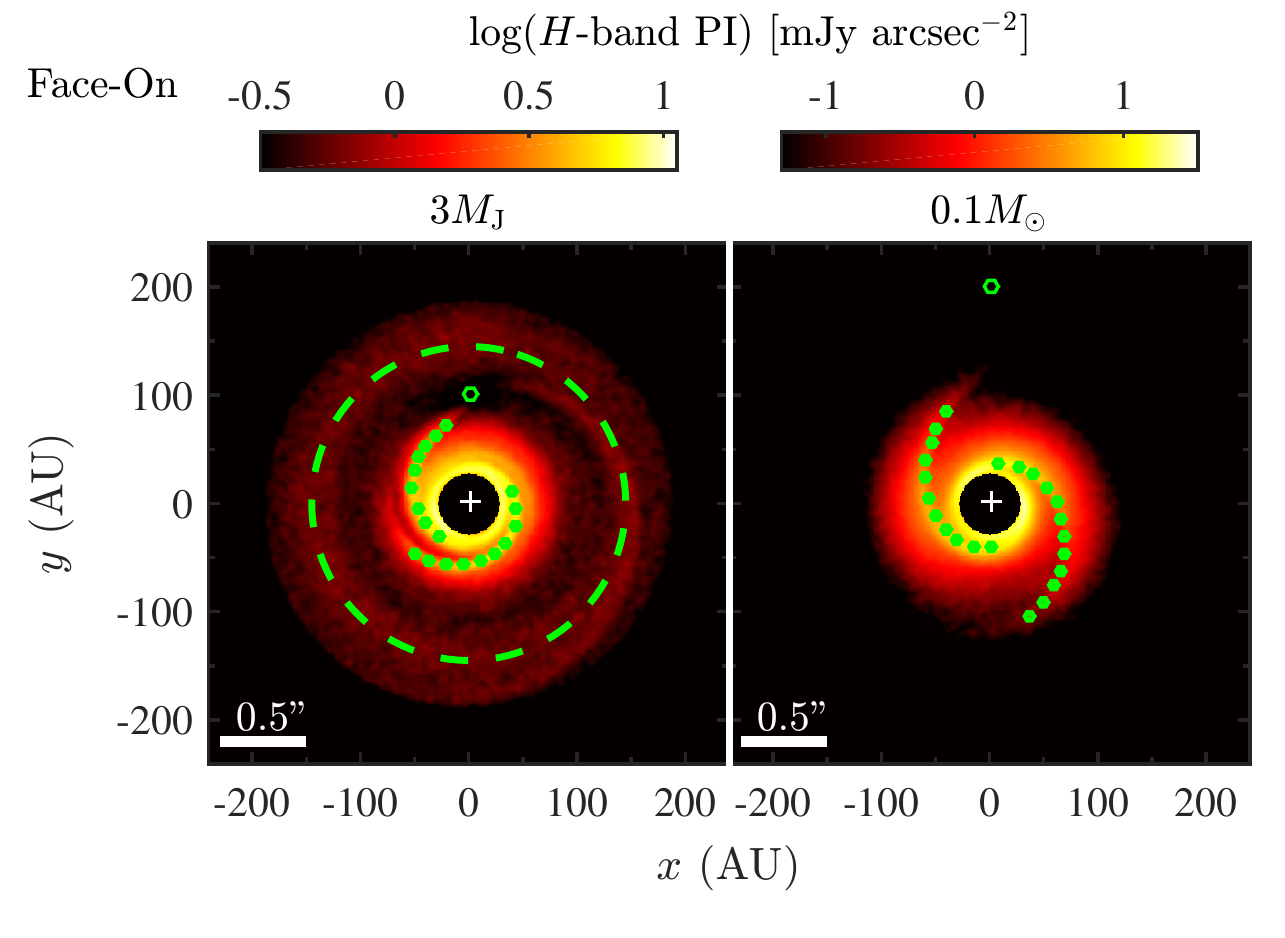}
\end{center}
\figcaption{Convolved polarized intensity $H$-band images for the two models seen face-on. The green dots trace out the locus of the two spiral arms in each case. The location of the companion in each panel is indicated by the open green circle. The dashed green circle in the $3\mj$ model marks the peak intensity in the outer disk.
\label{fig:image_locus}}
\end{figure}

\begin{figure}
\begin{center}
\includegraphics[trim=0 0 0 0, clip,width=\textwidth,angle=0]{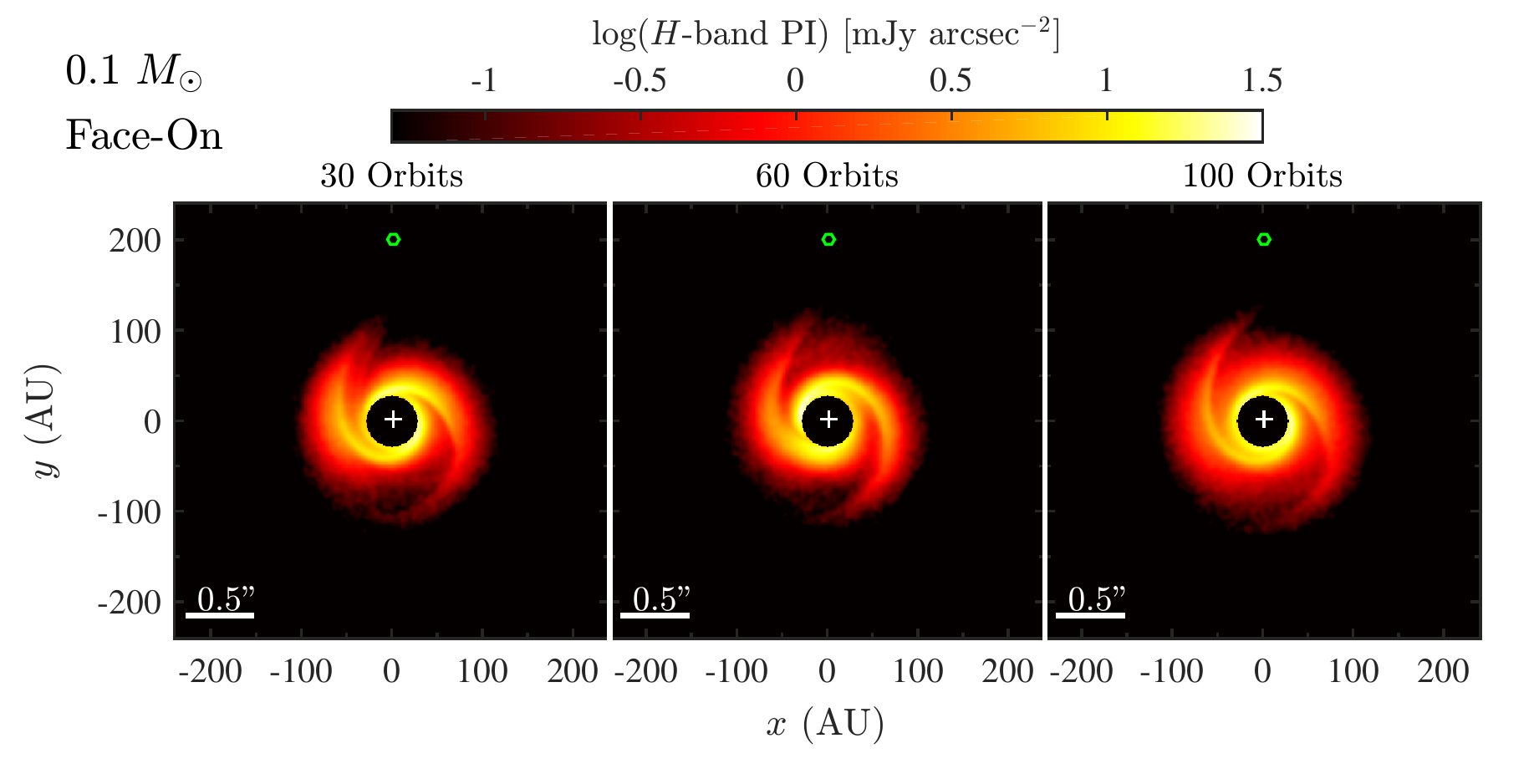}
\end{center}
\figcaption{Convolved polarized intensity $H$-band images for the 0.1~$M_\odot$ model at 30, 60, and 100 orbits. The green circle marks the position of the companion. The disk reaches a quasi-steady state after $\sim$10 orbits; the morphology of the arms stays roughly the same afterwards.
\label{fig:image_100mj_orbits}}
\end{figure}

\begin{figure}
\vspace*{-1cm}
\begin{center}
\includegraphics[trim=0 0 0 0, clip,width=\textwidth,angle=0]{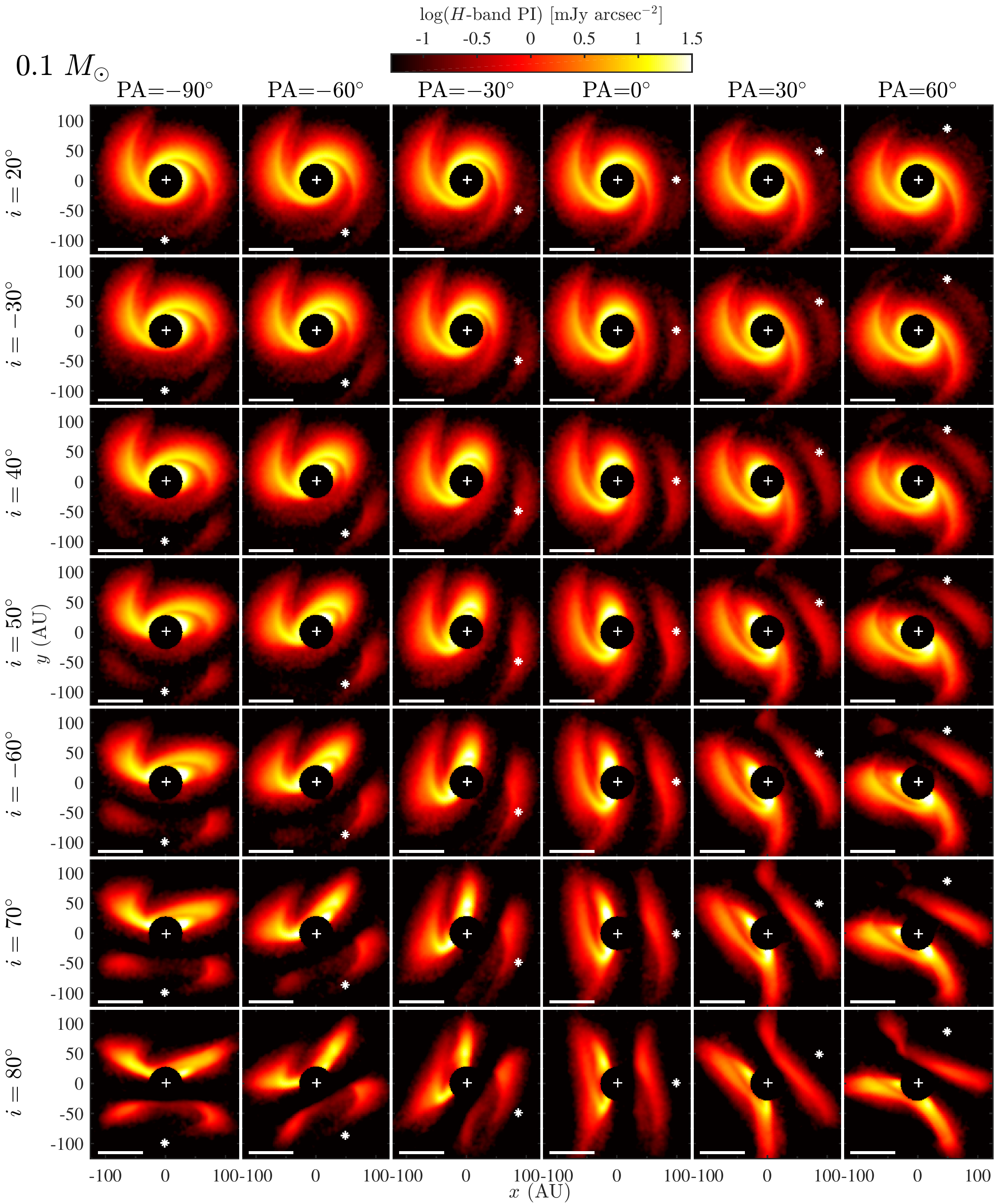}
\end{center}
\figcaption{Convolved $H$-band PI images for the $0.1M_\odot$ model seen at
position angles ranging from $-90^\circ$ to $60^\circ$ (PA runs from left to right) and inclinations from $20^\circ$ to $80^\circ$ ($i$ runs from top to bottom). The disk's nearside is marked by the white asterisk on the minor axis. The white bar at the lower left corner is $0.5\arcsec$ long. See Figure~\ref{fig:image_geometry} for  definitions of geometrical terms,
and Figure \ref{fig:image_100mj_geometry_big} for this same figure showing the projected
locations of the companion.
\label{fig:image_100mj_geometry}}
\end{figure}

\begin{figure}
\begin{center}
\includegraphics[trim=0 0 0 0, clip,width=\textwidth,angle=0]{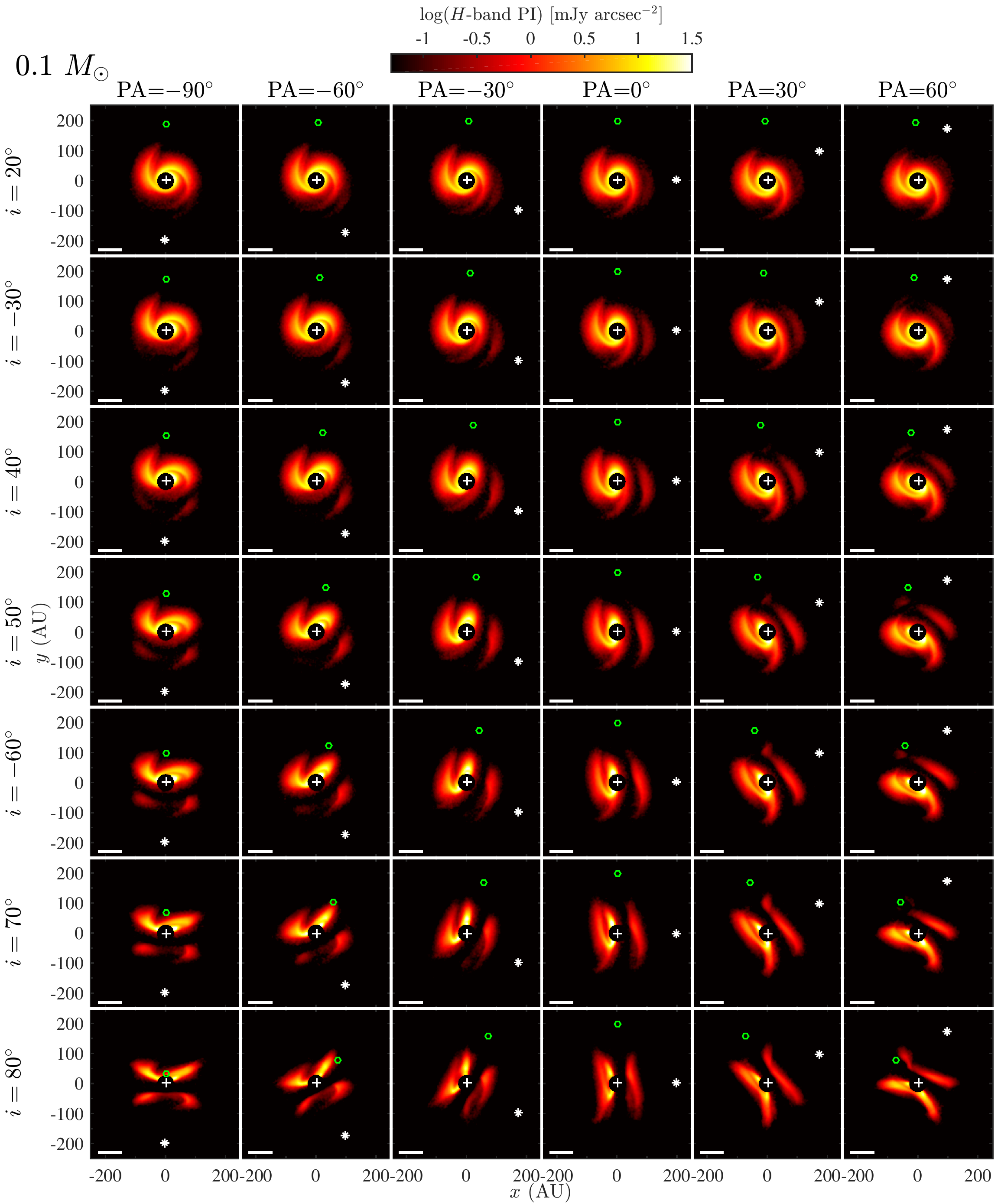}
\end{center}
\figcaption{Same as Figure~\ref{fig:image_100mj_geometry}, but zooming out to show the projected locations of the $0.1\msun$ companion (open green circle).
\label{fig:image_100mj_geometry_big}}
\end{figure}

In this paper, we carried out 3D hydrodynamics simulations of companion-induced spiral arms and gaps in protoplanetary disks. We examined the observational signatures of these features by synthesizing direct imaging observations at near-infrared wavelengths using 3D Monte Carlo radiative transfer simulations. We studied two models, one with a $3\mj$ planet at 100 AU and another with a $0.1\msun$ companion at 200 AU (assuming the central star is 1 $\msun$), surveying the entire parameter space in viewing inclination and position angle (Figures~\ref{fig:image_100mj_geometry} and \ref{fig:image_3mj_geometry}). In the $3\mj$ model the planet opens a gap and excites two prominent spiral arms in the inner disk; the inner disk arms, the outer disk ring, and the gap are readily detected when viewed face-on (Figure~\ref{fig:basic}). In the $0.1\msun$ model, the companion truncates the circumprimary disk to about half its orbital radius, and produces a prominent pair of arms with azimuthal wavenumber $m=2$ in face-on images (Figure~\ref{fig:basic}). Our results for face-on systems confirm previous findings in \citet{dong15spiralarms} and \citet{dong16hd100453} based on {\it Athena}++ (Stone et al., in preparation) hydro models, and can be compared with (and provide predictions for) current and future NIR imaging observations using Gemini/GPI, VLT/NACO/SPHERE, and Subaru/HiCIAO/SCExAO.

Below, we summarize the arm and gap morphologies presented in this paper:
\begin{enumerate}
\item Companion-induced spiral structures quickly settle into a quasi-steady state after $\sim$10 orbits (Figure~\ref{fig:image_100mj_orbits}). For face-on views, the morphologies are insensitive to observing wavelength and to the total ISM dust mass assumed, suggesting the positions of the arms are largely independent of altitude. Total intensity and polarized intensity images of face-on disks exhibit the same morphologies (Figures~\ref{fig:image_100mj_pianalysis} and \ref{fig:image_3mj_geometry_gap}). 
\item Depending on the viewing inclination and position angle, the trailing-double-arm pattern for face-on views may appear as one trailing arm; two trailing arms to one side of the star; or two arms on the same side but winding in opposite directions (e.g., Figure~\ref{fig:image_100mj_pianalysis}).
\item At inclinations $i\gtrsim50^\circ$, the bottom (obscured) half of the disk emerges. In the $3\mj$ model the nearside of the bottom half of the outer disk ring can be seen, and is always symmetric about the minor axis independent of the viewing angle (Figure~\ref{fig:image_3mj_geometry_gap}). In the $0.1\msun$ model, the two spirals on the bottom half are directly visible as they are not blocked by an outer disk; the bottom half in this case is  asymmetric to varying degrees depending on the position angle (Figures~\ref{fig:image_100mj_geometry} and \ref{fig:image_100mj_pianalysis}).
\item We confirmed the findings of \citet{canovas15qrur} that $Q_r$ (Equation~\ref{eq:qr}) and PI are generally similar. By comparison, $U_r$ (Equation~\ref{eq:ur}) represents observational noise in relatively face-on systems, but does contain signal for systems with $i\gtrsim40^\circ$ (Figure~\ref{fig:image_100mj_pianalysis}).
\end{enumerate}

We also highlight the shortcomings of two standard image reconstruction procedures:
\begin{enumerate}
\item The conventional practice of deprojecting inclined disk images based on the known inclination cannot restore the face-on morphology (Figures~\ref{fig:image_3mj_deproj} and \ref{fig:image_100mj_deproj}), particularly for high inclinations. The procedure (1) cannot recover the total number of arms; (2) cannot recover the arm locations; (3) cannot restore gaps and outer disk rings back to concentric circular structures; and (4) cannot remove the dark lane of the disk midplane. The dynamical planet-mass measurement technique based on arm morphology \citep{fung15} applies only to disks with inclinations $\lesssim20^\circ$.
\item The rescaling of images by $r^2$ introduces artificial azimuthal variations, for the obvious reason that in such systems, the physical distance to the star can differ from the projected distance.
\end{enumerate}

Finally, we make the following connections between our models and observations:
\begin{enumerate}
\item Asymmetric structures in nearly edge-on disks may be caused by spiral arms. Big and wide arms excited by massive (e.g., stellar mass) companions can generate global scale asymmetries, similar to those seen in HST observations of the HH~30 disk which is inclined by $84^\circ$ \citep{stapelfeldt99}. Narrower arms excited by planet-mass companions viewed from certain angles can collapse along the line of sight to appear as clumps in nearly edge-on disks (Figure~\ref{fig:image_3mj_edgeon}).
\item The partial gap at $r\sim85$~AU in AB Aur observed by Subaru/HiCIAO \citep{hashimoto11} may be a gap opened by a $3\mj$ planet at 67 AU. In this interpretation, the disk viewed at $20^\circ$ inclination and $-135^\circ$ position angle rotates counterclockwise on the sky, and has a nearside to the southeast, consistent with observations (Figure~\ref{fig:image_abaur}).
\item The single arm observed at $\sim$130~AU in the HD~141569~A disk \citep[][HST]{konishi16} may be a trailing-double-arm structure viewed at $\sim$$45^\circ$ inclination. 
\item At $i\gtrsim50^\circ$, the disk ring outside the gap opened by a giant planet manifests itself as two pseudo-arms placed symmetrically about the minor axis winding in opposite directions (Figure~\ref{fig:image_3mj_geometry_gap}). Our model image of a disk perturbed by a $3\mj$ planet at 25 AU, when viewed at $70^\circ$ inclination, resembles the AK Sco system as observed by SPHERE (\citealt{janson16}; Figure~\ref{fig:image_aksco}). A similar interpretation may explain the pair of minor-axis-symmetric arms at $r\sim40$~AU in the inner disk of HD 100546, as viewed at $\sim$$45^\circ$ inclination
(\citealt{currie15}, GPI; \citealt{garufi16}, SPHERE).
Our models make a prediction that can be tested by observations in the near future: these arms should have a much higher polarization fraction than the rest of the disk (see the two PF ``wings'' indicated by the arrows in Figure~\ref{fig:image_3mj_geometry_gap}). A deeper exposure may also reveal the dark lane of the disk midplane parallel to these pseudo-arms.
\end{enumerate}


\section*{Acknowledgments}
We thank Markus Janson and Jun Hashimoto for kindly sharing with us the SPHERE image of AK Sco and the HiCIAO image of AB Aur, respectively. An anonymous referee provided a helpful and encouraging report. This project is supported by NASA through Hubble Fellowship grant HST-HF-51320.01-A (R.D.) awarded by the Space Telescope Science Institute, which is operated by the Association of Universities for Research in Astronomy, Inc., for NASA, under contract NAS 5-26555. E.C. acknowledges support from NASA and the NSF. J.F. is grateful for the support from the Center for Integrative Planetary Science at the University of California, Berkeley. Numerical calculations were performed on the SAVIO cluster provided by the Berkeley Research Computing program, supported by the UC Berkeley Vice Chancellor for Research and the Berkeley Center for Integrative Planetary Science.


\clearpage


\begin{figure}
\vspace*{-0.7cm}
\begin{center}
\includegraphics[trim=0 0 0 0, clip,width=\textwidth,angle=0]{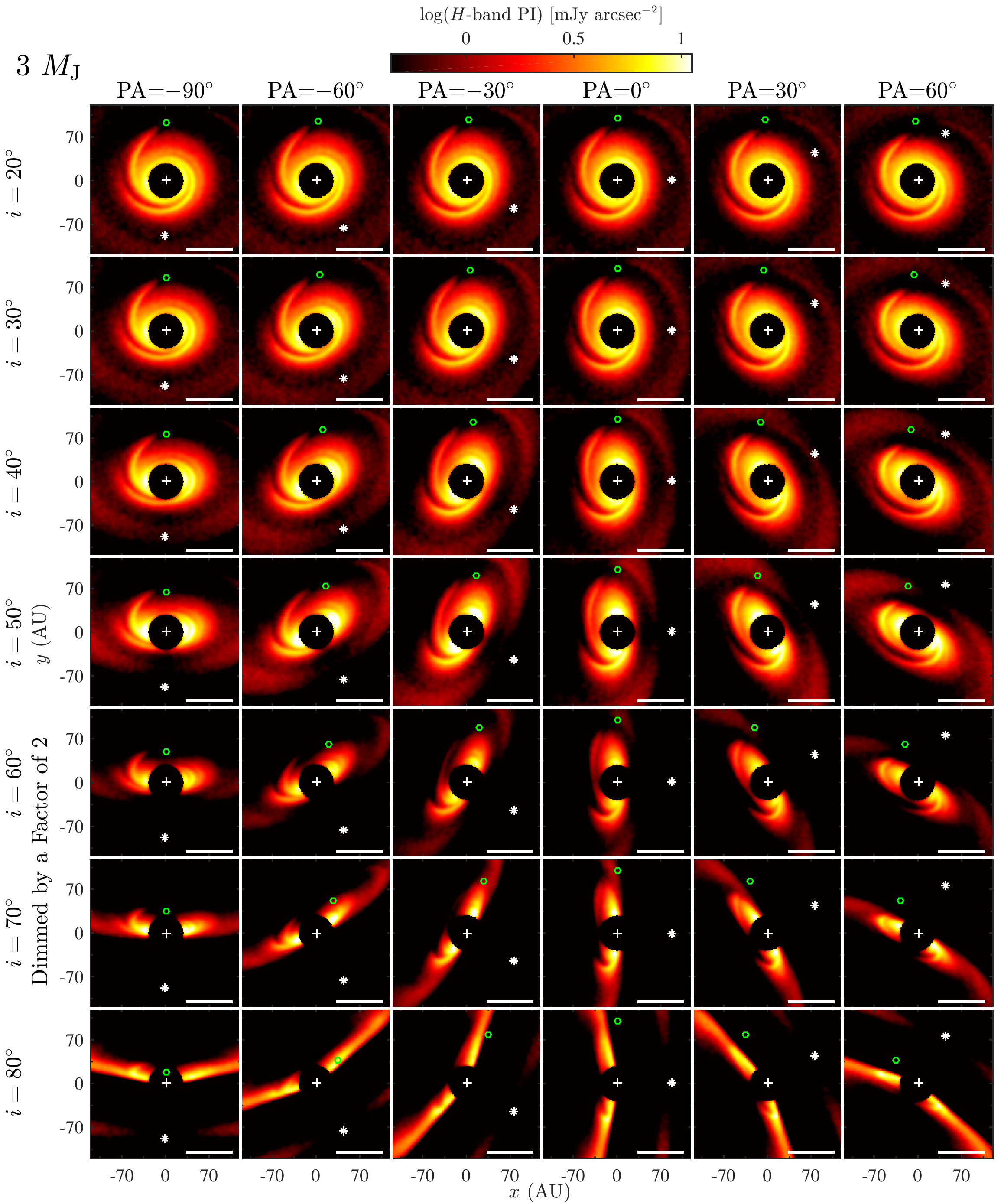}
\end{center}
\figcaption{Same as Figure~\ref{fig:image_100mj_geometry}, but for the $3\mj$ model. See Figure~\ref{fig:image_geometry} for definitions of angles and geometrical terms. %
Each panel is 240 AU on a side to highlight the inner spiral arms. The open green circle indicates the projected location of the $3\mj$ planet. The panels at $60^\circ$ and $70^\circ$ inclinations are dimmed by a factor of 2 to fit within the color scaling.
}
\end{figure}

\begin{figure}
  \ContinuedFloat 
\begin{center}
\includegraphics[trim=0 0 0 0, clip,width=\textwidth,angle=0]{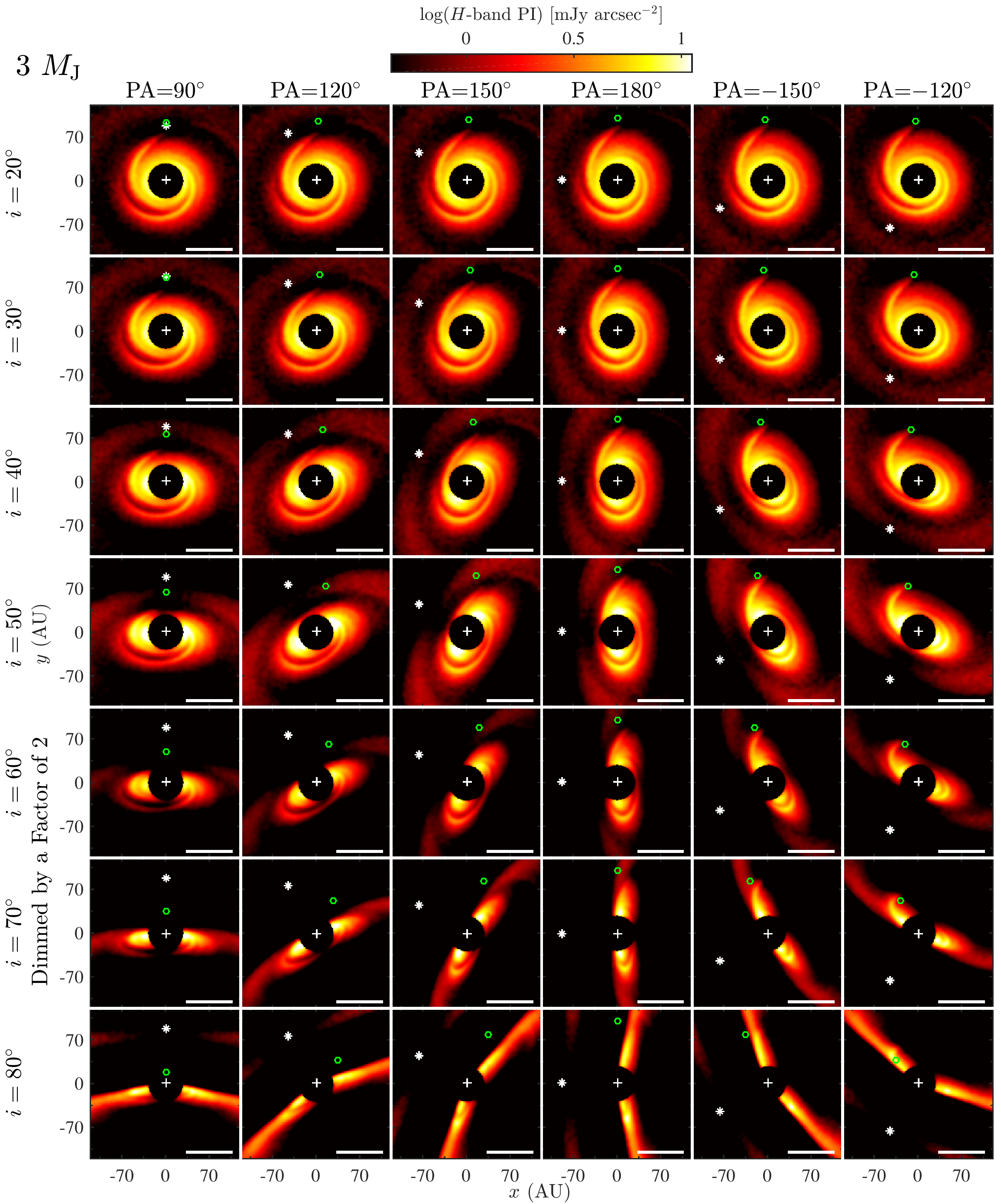}
\end{center}
\figcaption{Continuation of the $3M_{\rm J}$ model results for more position angles.
\label{fig:image_3mj_geometry}}
\end{figure}

\begin{figure}
\vspace*{-0.7cm}
\begin{center}
\includegraphics[trim=0 0 0 0, clip,width=\textwidth,angle=0]{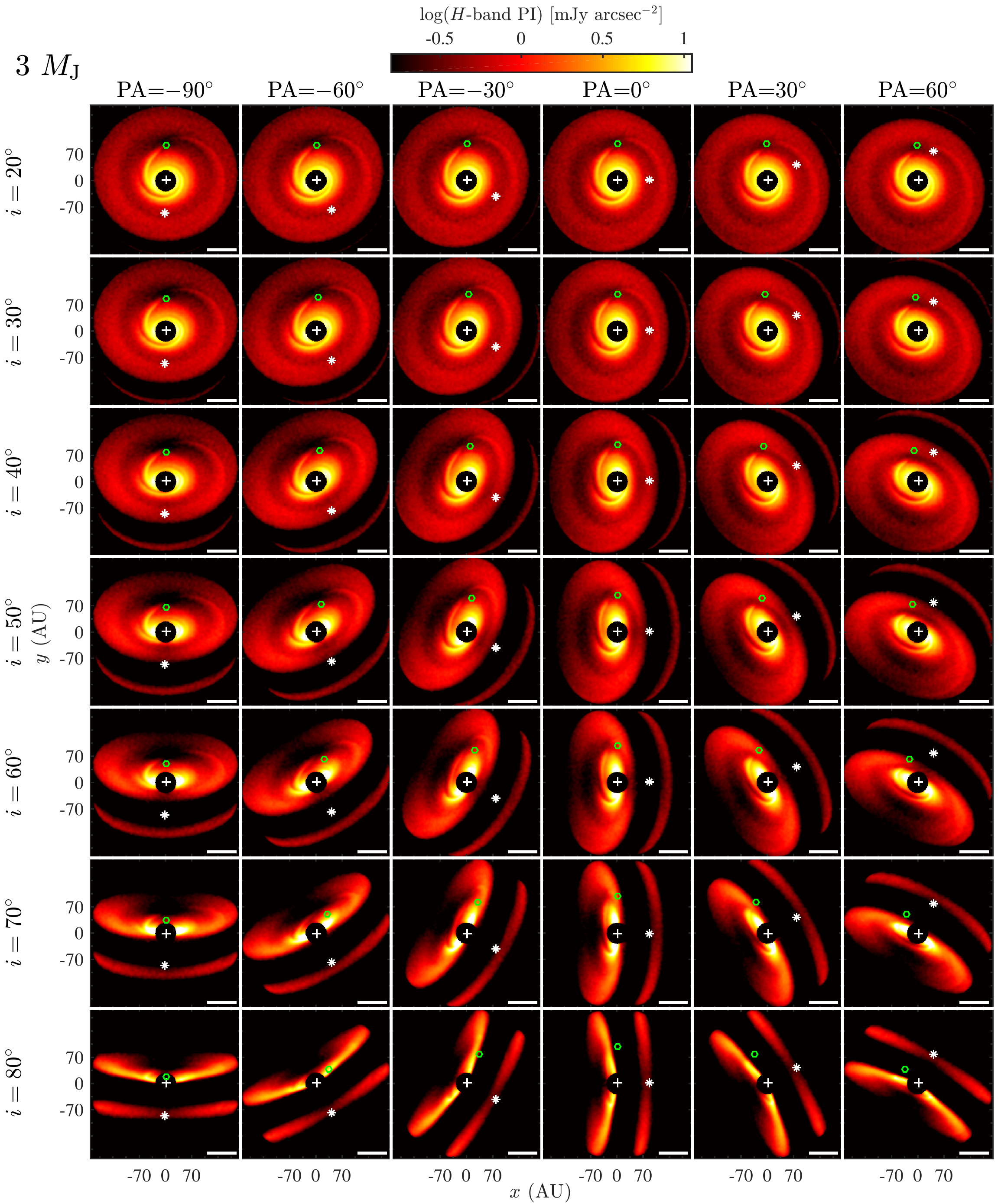}
\end{center}
\figcaption{Same as Figure~\ref{fig:image_3mj_geometry}, but each panel is 400 AU on a side to highlight the gap, the outer disk, and the bottom half of the disk.
}
\end{figure}

\begin{figure}
  \ContinuedFloat 
\begin{center}
\includegraphics[trim=0 0 0 0, clip,width=\textwidth,angle=0]{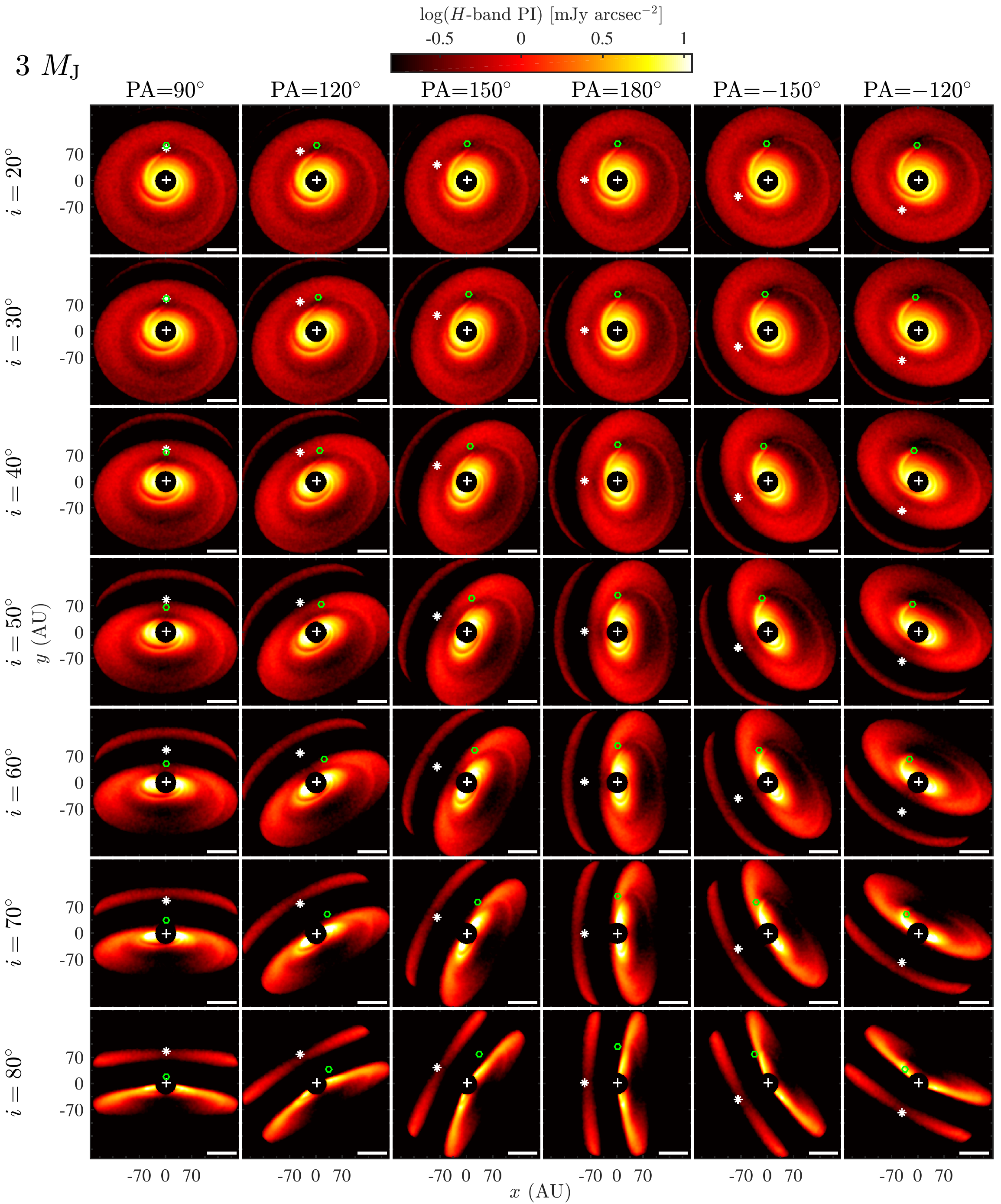}
\end{center}
\figcaption{Continuation of the $3M_{\rm J}$ model results for more position angles.
\label{fig:image_3mj_geometry_2}}
\end{figure}

\begin{figure}
\begin{center}
\includegraphics[trim=0 0 0 0, clip,width=\textwidth,angle=0]{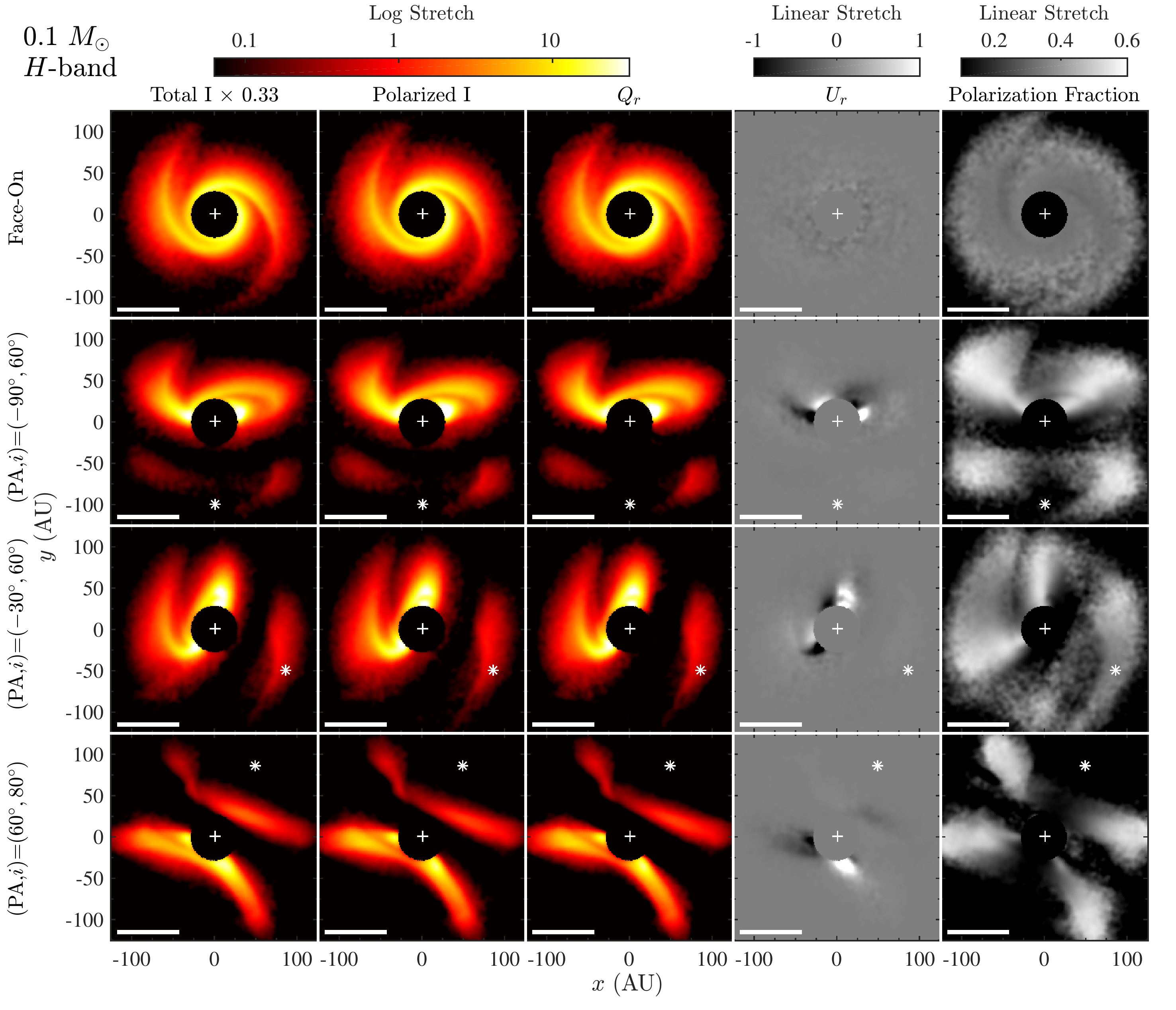}
\end{center}
\figcaption{From left to right: total intensity (TI) divided by 3 (to fit within the color scaling), polarized intensity (PI), $Q_r$ (Equation~\ref{eq:qr}), $U_r$ (Equation~\ref{eq:ur}), and polarization fraction PI/TI, for a face-on view and three other inclined views (as labeled to the left) for the $0.1M_\odot$ model. See Figure~\ref{fig:image_geometry} for the definitions of geometrical terms. The first three columns are in log stretch while the last two are in linear stretch. The white bar at the lower left corner indicates $0.5\arcsec$. Depending on the viewing geometry, the two trailing arms seen on opposite sides of the face-on disk (top row) may appear as two trailing arms on one side of the disk (second row), one trailing arm (third row), or two arms to one side winding in opposite directions (bottom row). $Q_r$, PI, and TI resemble each other in all cases. $U_r$ has no signal and represents the noise level in the face-on case, but contains some signal in the inner disk in inclined systems. See Section~\ref{sec:results_arms} for details.
\label{fig:image_100mj_pianalysis}}
\end{figure}

\begin{figure}
\begin{center}
\includegraphics[trim=0 0 0 0, clip,width=0.65\textwidth,angle=0]{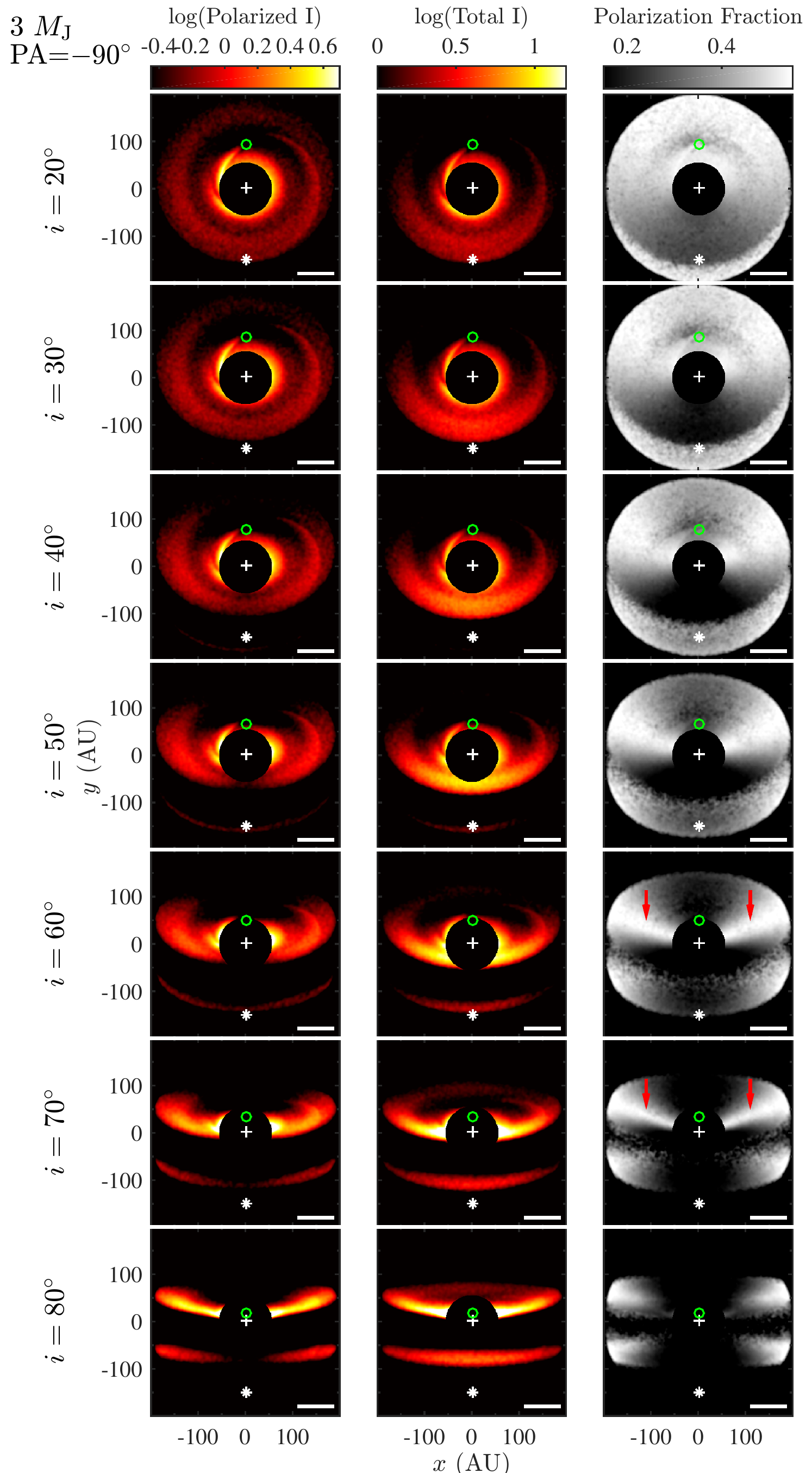}
\end{center}
\figcaption{\footnotesize{Polarized intensity (left), total intensity (middle), and polarization fraction (PI/TI; right) for the $3\mj$ model at PA=90$^\circ$ and various inclinations (labeled on the left). See Figure~\ref{fig:image_geometry} for  definitions of geometrical terms. The white bar at the lower right corner indicates $0.5\arcsec$. The inner $0.4\arcsec$ (56 AU) in radius is blocked and the panels are 400 AU on each side to highlight the gap and the outer disk ring. The nearside of the disk is marked by the white asterisk, and the projected location of the $3\mj$ planet is marked by the green circle. The red arrows in the right column indicate the high PF wings. See Section~\ref{sec:results_gap} for details.}
\label{fig:image_3mj_geometry_gap}}
\end{figure}

\begin{figure}
\begin{center}
\includegraphics[trim=0 0 0 0, clip,width=\textwidth,angle=0]{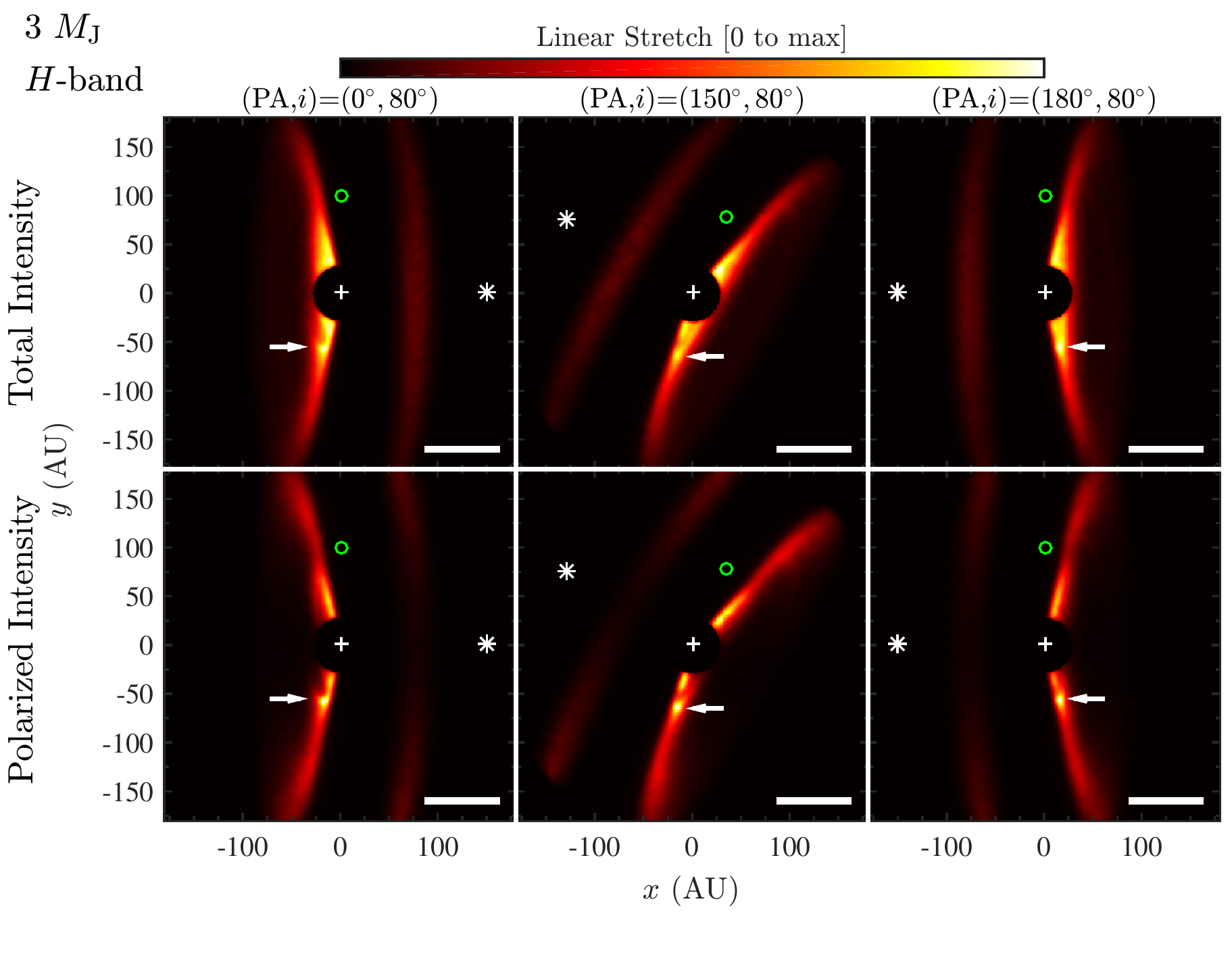}
\end{center}
\figcaption{TI (top) and PI (bottom) images of the $3\mj$ model in linear stretch (0 to maximum) at three position angles with $i=80^\circ$ (as annotated on the top of the figure). See Figure~\ref{fig:image_geometry} for definitions of geometrical terms. The bar at the lower right corner indicates $0.5\arcsec$. The open green circle indicates the projected location of the $3\mj$ planet. The white arrows point to  clumps that are spiral arms compressed along the line of sight in these disks viewed nearly edge-on. See Section~\ref{sec:results_gap} for details.
\label{fig:image_3mj_edgeon}}
\end{figure}

\begin{figure}
\begin{center}
\includegraphics[trim=0 0 0 0, clip,width=0.85\textwidth,angle=0]{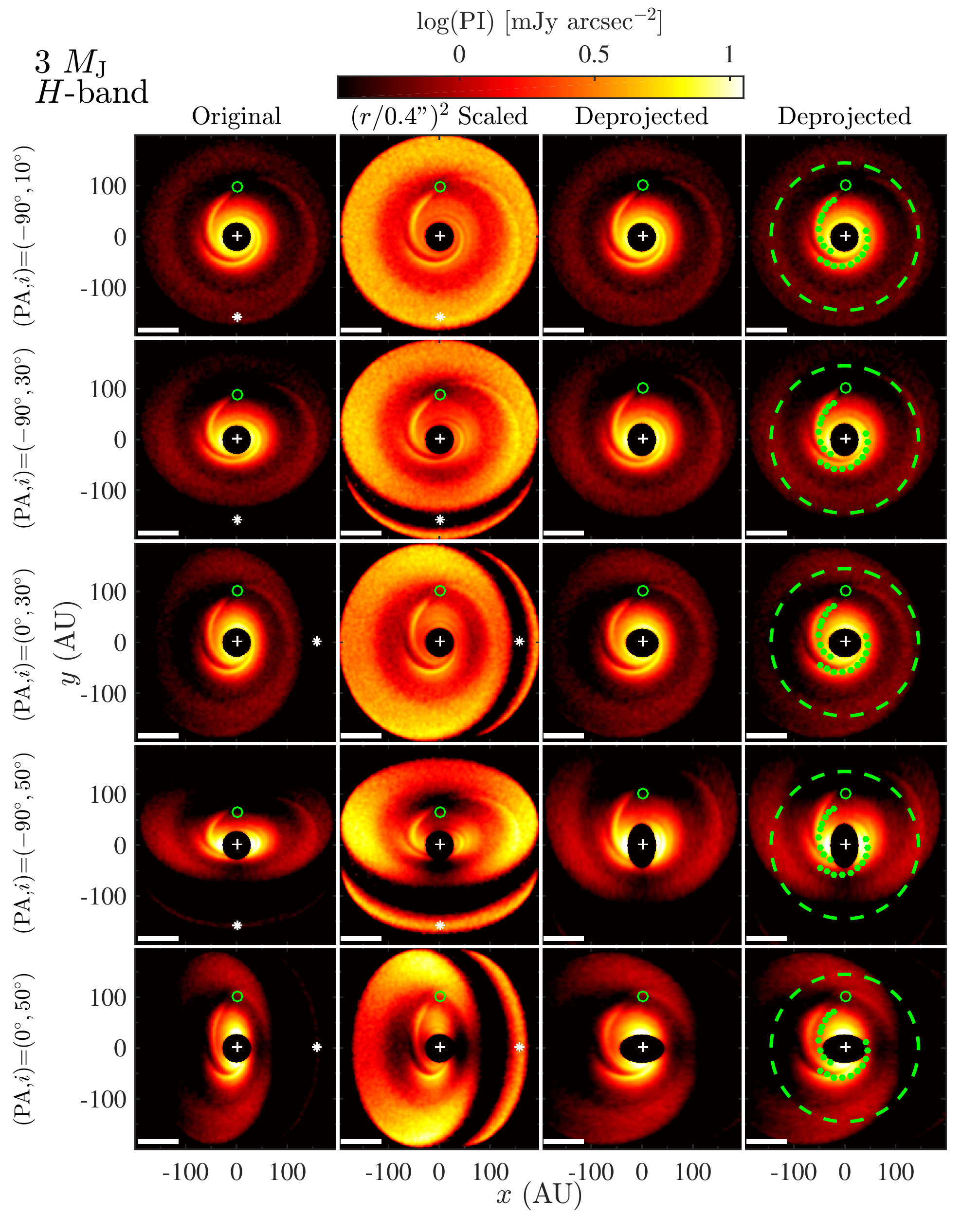}
\end{center}
\figcaption{\footnotesize{Artifacts from $r^2$-scaling and deprojecting images. The left column is the original convolved PI images at four viewing angles as indicated on the left. The second column contains the original images scaled by $r^2$. The third column contains the original images deprojected by their inclinations. The last column is the same as the third but with the locus of the two arms and the location of the outer disk ring in face-on images (Figure~\ref{fig:image_locus}) overlaid.  The bar at the lower right corner indicates $0.5\arcsec$.
The open green circle indicates the projected location of the $3\mj$ planet in the first two columns and the deprojected location of the planet (i.e., 100 AU to the North) in the last two columns. The $r^2$-scaling procedure enhances features at large angular distances, but also introduces artificial azimuthal variations. Deprojection cannot restore the locations of the arms viewed face-on, nor can it restore the off-centered elliptical gap and outer disk ring back to centered circular structures.}
\label{fig:image_3mj_deproj}}
\end{figure}

\begin{figure}
\begin{center}
\includegraphics[trim=0 0 0 0, clip,width=0.9\textwidth,angle=0]{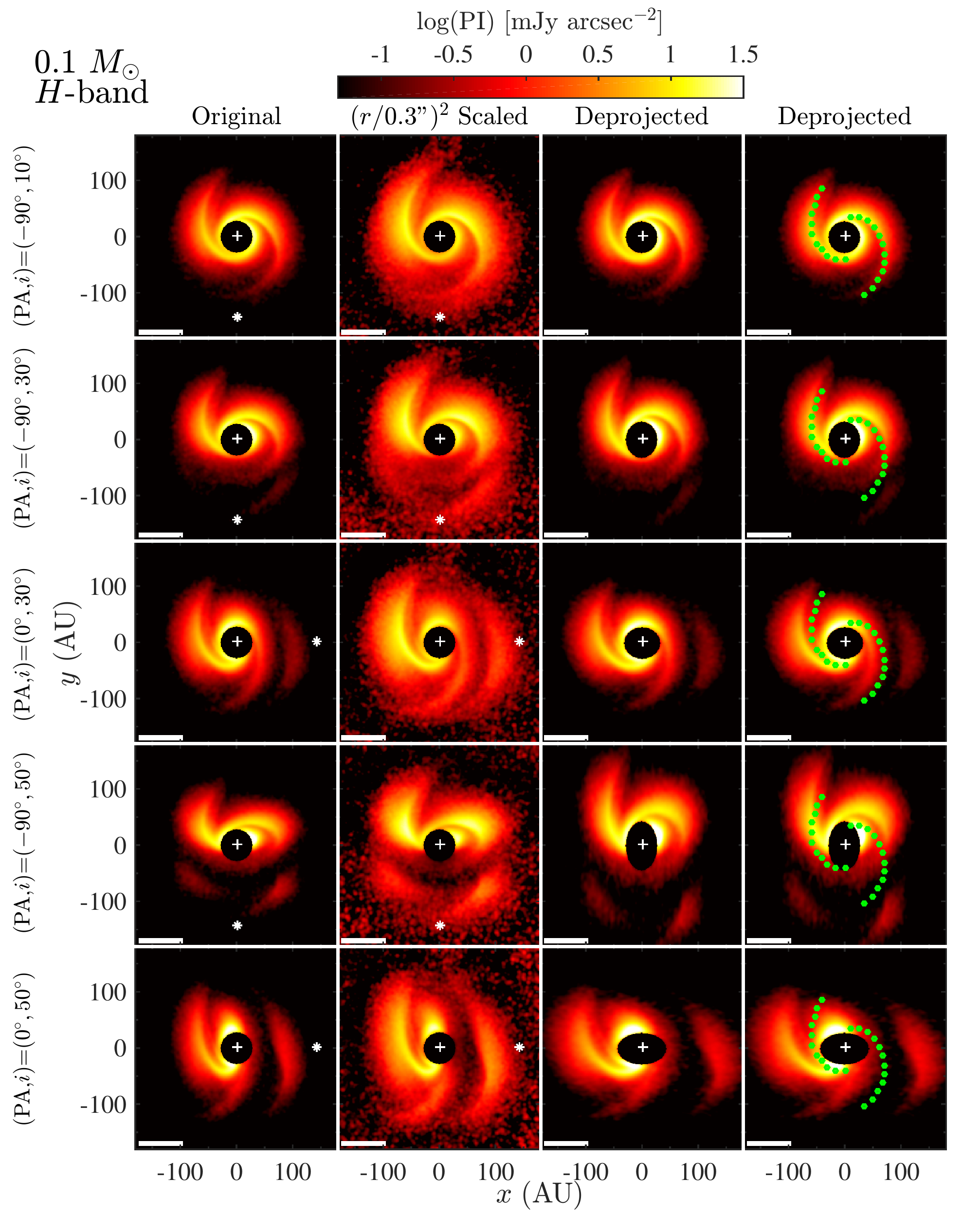}
\end{center}
\figcaption{Same as Figure~\ref{fig:image_3mj_deproj}, but for the $0.1M_\odot$ model. Deprojection does not restore the face-on morphology of the arms, nor does it remove the dark lane at the disk midplane and the nearside of the disk's bottom (obscured) half as seen in inclined disks.
\label{fig:image_100mj_deproj}}
\end{figure}

\begin{figure}
\begin{center}
\includegraphics[trim=0 0 0 0, clip,width=\textwidth,angle=0]{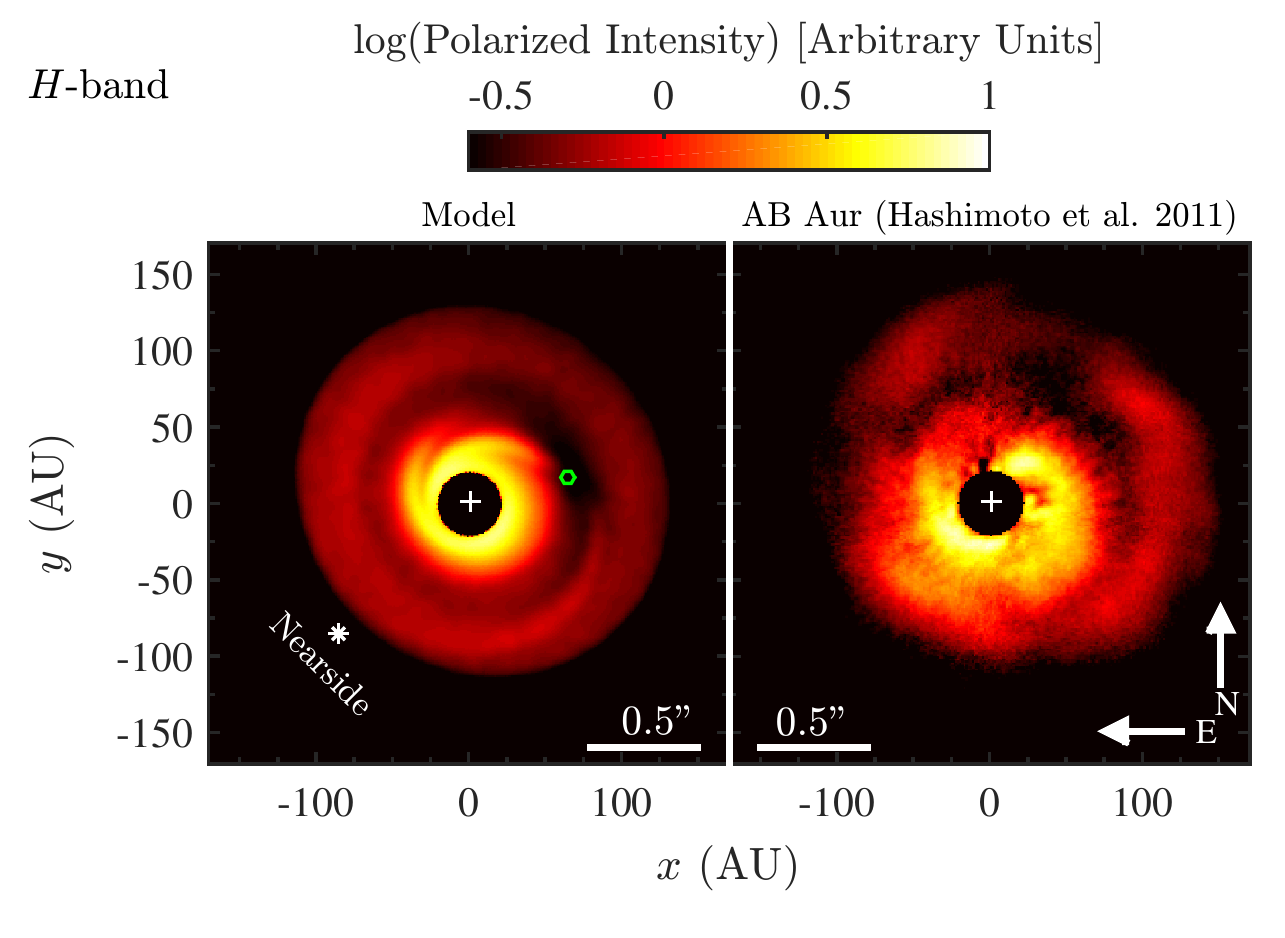}
\end{center}
\figcaption{Synthetic $3\mj$ model image (left) and Subaru/HiCIAO observation of AB Aur \citep[right;][]{hashimoto11} in $H$-band polarized intensity. Our model has been shrunk by a factor of 1.5 to match the physical size of the observed disk (thus the planet is now at 67 AU; its projected location is indicated by the open green circle). The inner circle of radius $r=0.15\arcsec$ is blocked to simulate
a coronograph. The angular resolution is $0.06\arcsec$ in both images. The geometry of the model is consistent with observations: the disk is viewed at 
PA$=-135^\circ$ and $i=20^\circ$, rotates counterclockwise on the sky, and has a  nearside to the southeast \citep{fukagawa04, hashimoto11, tang12}. Viewed at an inclination of $20^\circ$, the gap opened by the planet in the model resembles the observed gap.
\label{fig:image_abaur}}
\end{figure}

\begin{figure}
\begin{center}
\includegraphics[trim=0 0 0 0, clip,width=\textwidth,angle=0]{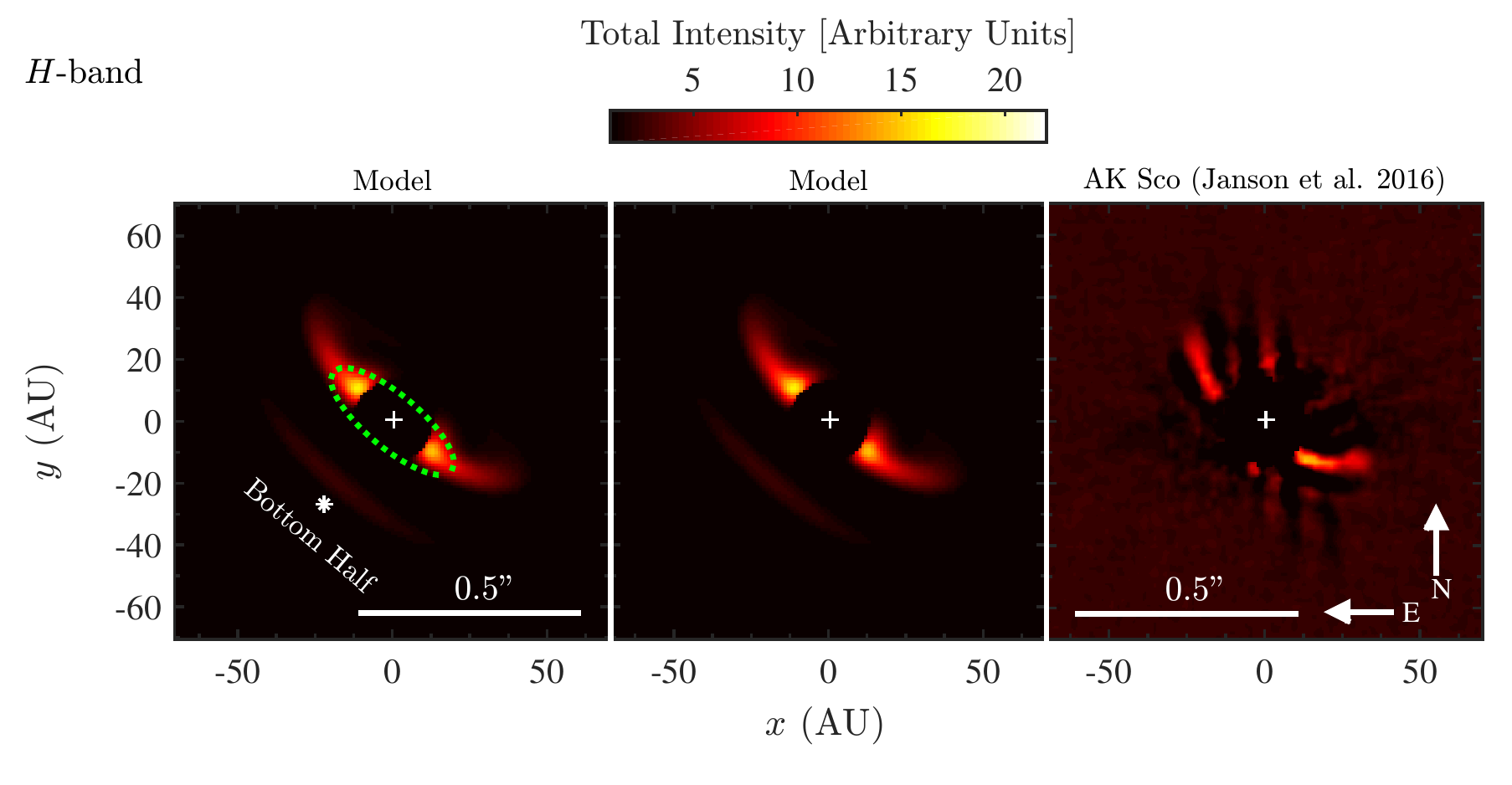}
\end{center}
\figcaption{Synthetic $3\mj$ model image (left and middle; identical except labels) and the VLT/SPHERE observation of AK Sco \citep[right;][]{janson16} in $H$-band total intensity. Our model has been shrunk by a factor of 4 to match the physical size of the observed disk (thus the planet is now at 25 AU). 
The inner circle of radius $r=0.0925\arcsec$ is blocked off to simulate a coronagraph. The angular resolution is $0.04\arcsec$ in both images. The model image is viewed at a position angle of $-130^\circ$ and an inclination of $70^\circ$; these same parameters characterize AK Sco as constrained by CO observations \citep{czekala15}. The bottom (obscured) half of the model disk is labeled, and the planet's orbit (circular and coplanar with the disk) projected on the sky is marked by the green dotted circle. The apparent spiral-arm-like features (``pseudo-arms'') in the model actually arise from the outer disk ring viewed at high inclination (note their morphologies are insensitive to the specific location of the planet in the azimuthal direction).
The pseudo-arms in the observations appear to be sharper, possibly because of the angular differential imaging (ADI) data reduction process.
\label{fig:image_aksco}}
\end{figure}

\clearpage\end{CJK*}

\begin{thebibliography}{}
\expandafter\ifx\csname natexlab\endcsname\relax\def\natexlab#1{#1}\fi

\bibitem[{{Avenhaus} {et~al.}(2014){Avenhaus}, {Quanz}, {Meyer}, {Brittain},
  {Carr}, \& {Najita}}]{avenhaus14hd100546}
{Avenhaus}, H., {Quanz}, S.~P., {Meyer}, M.~R., {Brittain}, S.~D., {Carr},
  J.~S., \& {Najita}, J.~R. 2014, \apj, 790, 56

\bibitem[{{Benisty} {et~al.}(2015){Benisty}, {Juhasz}, {Boccaletti},
  {Avenhaus}, {Milli}, {Thalmann}, {Dominik}, {Pinilla}, {Buenzli}, {Pohl},
  {Beuzit}, {Birnstiel}, {de Boer}, {Bonnefoy}, {Chauvin}, {Christiaens},
  {Garufi}, {Grady}, {Henning}, {Huelamo}, {Isella}, {Langlois}, {M{\'e}nard},
  {Mouillet}, {Olofsson}, {Pantin}, {Pinte}, \& {Pueyo}}]{benisty15}
{Benisty}, M., {et~al.} 2015, \aap, 578, L6

\bibitem[{{Beuzit} {et~al.}(2008){Beuzit}, {Feldt}, {Dohlen}, {Mouillet},
  {Puget}, {Wildi}, {Abe}, {Antichi}, {Baruffolo}, {Baudoz}, {Boccaletti},
  {Carbillet}, {Charton}, {Claudi}, {Downing}, {Fabron}, {Feautrier},
  {Fedrigo}, {Fusco}, {Gach}, {Gratton}, {Henning}, {Hubin}, {Joos}, {Kasper},
  {Langlois}, {Lenzen}, {Moutou}, {Pavlov}, {Petit}, {Pragt}, {Rabou}, {Rigal},
  {Roelfsema}, {Rousset}, {Saisse}, {Schmid}, {Stadler}, {Thalmann}, {Turatto},
  {Udry}, {Vakili}, \& {Waters}}]{beuzit08}
{Beuzit}, J.-L., {et~al.} 2008, in Society of Photo-Optical Instrumentation
  Engineers (SPIE) Conference Series, Vol. 7014, Society of Photo-Optical
  Instrumentation Engineers (SPIE) Conference Series

\bibitem[{{Boccaletti} {et~al.}(2013){Boccaletti}, {Pantin}, {Lagrange},
  {Augereau}, {Meheut}, \& {Quanz}}]{boccaletti13}
{Boccaletti}, A., {Pantin}, E., {Lagrange}, A.-M., {Augereau}, J.-C., {Meheut},
  H., \& {Quanz}, S.~P. 2013, \aap, 560, A20

\bibitem[{{Boccaletti} {et~al.}(2015){Boccaletti}, {Thalmann}, {Lagrange},
  {Janson}, {Augereau}, {Schneider}, {Milli}, {Grady}, {Debes}, {Langlois},
  {Mouillet}, {Henning}, {Dominik}, {Maire}, {Beuzit}, {Carson}, {Dohlen},
  {Engler}, {Feldt}, {Fusco}, {Ginski}, {Girard}, {Hines}, {Kasper}, {Mawet},
  {M{\'e}nard}, {Meyer}, {Moutou}, {Olofsson}, {Rodigas}, {Sauvage},
  {Schlieder}, {Schmid}, {Turatto}, {Udry}, {Vakili}, {Vigan}, {Wahhaj}, \&
  {Wisniewski}}]{boccaletti15}
{Boccaletti}, A., {et~al.} 2015, \nat, 526, 230

\bibitem[{{Canovas} {et~al.}(2015){Canovas}, {M{\'e}nard}, {de Boer}, {Pinte},
  {Avenhaus}, \& {Schreiber}}]{canovas15qrur}
{Canovas}, H., {M{\'e}nard}, F., {de Boer}, J., {Pinte}, C., {Avenhaus}, H., \&
  {Schreiber}, M.~R. 2015, \aap, 582, L7

\bibitem[{{Canovas} {et~al.}(2013){Canovas}, {M{\'e}nard}, {Hales},
  {Jord{\'a}n}, {Schreiber}, {Casassus}, {Gledhill}, \& {Pinte}}]{canovas13}
{Canovas}, H., {M{\'e}nard}, F., {Hales}, A., {Jord{\'a}n}, A., {Schreiber},
  M.~R., {Casassus}, S., {Gledhill}, T.~M., \& {Pinte}, C. 2013, \aap, 556,
  A123

\bibitem[{{Currie} {et~al.}(2015){Currie}, {Cloutier}, {Brittain}, {Grady},
  {Burrows}, {Muto}, {Kenyon}, \& {Kuchner}}]{currie15}
{Currie}, T., {Cloutier}, R., {Brittain}, S., {Grady}, C., {Burrows}, A.,
  {Muto}, T., {Kenyon}, S.~J., \& {Kuchner}, M.~J. 2015, ArXiv e-prints

\bibitem[{{Czekala} {et~al.}(2015){Czekala}, {Andrews}, {Jensen}, {Stassun},
  {Torres}, \& {Wilner}}]{czekala15}
{Czekala}, I., {Andrews}, S.~M., {Jensen}, E.~L.~N., {Stassun}, K.~G.,
  {Torres}, G., \& {Wilner}, D.~J. 2015, \apj, 806, 154

\bibitem[{{de Juan Ovelar} {et~al.}(2013){de Juan Ovelar}, {Min}, {Dominik},
  {Thalmann}, {Pinilla}, {Benisty}, \& {Birnstiel}}]{dejuanovelar13}
{de Juan Ovelar}, M., {Min}, M., {Dominik}, C., {Thalmann}, C., {Pinilla}, P.,
  {Benisty}, M., \& {Birnstiel}, T. 2013, \aap, 560, A111

\bibitem[{{Dong} {et~al.}(2012){Dong}, {Rafikov}, {Zhu}, {Hartmann}, {Whitney},
  {Brandt}, {Muto}, {Hashimoto}, {Grady}, {Follette}, {Kuzuhara}, {Tanii},
  {Itoh}, {Thalmann}, {Wisniewski}, {Mayama}, {Janson}, {Abe}, {Brandner},
  {Carson}, {Egner}, {Feldt}, {Goto}, {Guyon}, {Hayano}, {Hayashi}, {Hayashi},
  {Henning}, {Hodapp}, {Honda}, {Inutsuka}, {Ishii}, {Iye}, {Kandori}, {Knapp},
  {Kudo}, {Kusakabe}, {Matsuo}, {McElwain}, {Miyama}, {Morino}, {Moro-Martin},
  {Nishimura}, {Pyo}, {Suto}, {Suzuki}, {Takami}, {Takato}, {Terada}, {Tomono},
  {Turner}, {Watanabe}, {Yamada}, {Takami}, {Usuda}, \&
  {Tamura}}]{dong12cavity}
{Dong}, R., {et~al.} 2012, \apj, 750, 161

\bibitem[{{Dong} {et~al.}(2016){Dong}, {Zhu}, {Fung}, {Rafikov}, {Chiang}, \&
  {Wagner}}]{dong16hd100453}
{Dong}, R., {Zhu}, Z., {Fung}, J., {Rafikov}, R., {Chiang}, E., \& {Wagner}, K.
  2016, \apjl, 816, L12

\bibitem[{{Dong} {et~al.}(2015{\natexlab{a}}){Dong}, {Zhu}, {Rafikov}, \&
  {Stone}}]{dong15spiralarms}
{Dong}, R., {Zhu}, Z., {Rafikov}, R.~R., \& {Stone}, J.~M. 2015{\natexlab{a}},
  \apjl, 809, L5

\bibitem[{{Dong} {et~al.}(2015{\natexlab{b}}){Dong}, {Zhu}, \&
  {Whitney}}]{dong15gaps}
{Dong}, R., {Zhu}, Z., \& {Whitney}, B. 2015{\natexlab{b}}, \apj, 809, 93

\bibitem[{{Duffell}(2015)}]{duffell15gap}
{Duffell}, P.~C. 2015, \apjl, 807, L11

\bibitem[{{Duffell} \& {Dong}(2015)}]{duffell15dong}
{Duffell}, P.~C., \& {Dong}, R. 2015, \apj, 802, 42

\bibitem[{{Espaillat} {et~al.}(2014){Espaillat}, {Muzerolle}, {Najita},
  {Andrews}, {Zhu}, {Calvet}, {Kraus}, {Hashimoto}, {Kraus}, \&
  {D'Alessio}}]{espaillat14}
{Espaillat}, C., {et~al.} 2014, ArXiv e-prints

\bibitem[{{Follette} {et~al.}(2013){Follette}, {Tamura}, {Hashimoto},
  {Whitney}, {Grady}, {Close}, {Andrews}, {Kwon}, {Wisniewski}, {Brandt},
  {Mayama}, {Kandori}, {Dong}, {Abe}, {Brandner}, {Carson}, {Currie}, {Egner},
  {Feldt}, {Goto}, {Guyon}, {Hayano}, {Hayashi}, {Hayashi}, {Henning},
  {Hodapp}, {Ishii}, {Iye}, {Janson}, {Knapp}, {Kudo}, {Kusakabe}, {Kuzuhara},
  {McElwain}, {Matsuo}, {Miyama}, {Morino}, {Moro-Martin}, {Nishimura}, {Pyo},
  {Serabyn}, {Suto}, {Suzuki}, {Takami}, {Takato}, {Terada}, {Thalmann},
  {Tomono}, {Turner}, {Watanabe}, {Yamada}, {Takami}, \& {Usuda}}]{follette13}
{Follette}, K.~B., {et~al.} 2013, \apj, 767, 10

\bibitem[{{Fukagawa} {et~al.}(2004){Fukagawa}, {Hayashi}, {Tamura}, {Itoh},
  {Hayashi}, {Oasa}, {Takeuchi}, {Morino}, {Murakawa}, {Oya}, {Yamashita},
  {Suto}, {Mayama}, {Naoi}, {Ishii}, {Pyo}, {Nishikawa}, {Takato}, {Usuda},
  {Ando}, {Iye}, {Miyama}, \& {Kaifu}}]{fukagawa04}
{Fukagawa}, M., {et~al.} 2004, \apjl, 605, L53

\bibitem[{{Fung}(2015)}]{fung15thesis}
{Fung}, J. 2015, PhD thesis, University of Toronto, Canada

\bibitem[{{Fung} \& {Dong}(2015)}]{fung15}
{Fung}, J., \& {Dong}, R. 2015, \apjl, 815, L21

\bibitem[{{Fung} {et~al.}(2014){Fung}, {Shi}, \& {Chiang}}]{fung14}
{Fung}, J., {Shi}, J.-M., \& {Chiang}, E. 2014, \apj, 782, 88

\bibitem[{{Garufi} {et~al.}(2013){Garufi}, {Quanz}, {Avenhaus}, {Buenzli},
  {Dominik}, {Meru}, {Meyer}, {Pinilla}, {Schmid}, \& {Wolf}}]{garufi13}
{Garufi}, A., {et~al.} 2013, \aap, 560, A105

\bibitem[{{Garufi} {et~al.}(2016){Garufi}, {Quanz}, {Schmid}, {Mulders},
  {Avenhaus}, {Boccaletti}, {Ginski}, {Langlois}, {Stolker}, {Augereau},
  {Benisty}, {Lopez}, {Dominik}, {Gratton}, {Henning}, {Janson}, {Menard},
  {Meyer}, {Pinte}, {Sissa}, {Vigan}, {Zurlo}, {Bazzon}, {Buenzli}, {Bonnefoy},
  {Brandner}, {Chauvin}, {Cheetham}, {Cudel}, {Desidera}, {Feldt}, {Galicher},
  {Kasper}, {Lagrange}, {Lannier}, {Maire}, {Mesa}, {Mouillet}, {Peretti},
  {Perrot}, {Salter}, \& {Wildi}}]{garufi16}
---. 2016, ArXiv e-prints

\bibitem[{{Grady} {et~al.}(2013){Grady}, {Muto}, {Hashimoto}, {Fukagawa},
  {Currie}, {Biller}, {Thalmann}, {Sitko}, {Russell}, {Wisniewski}, {Dong},
  {Kwon}, {Sai}, {Hornbeck}, {Schneider}, {Hines}, {Moro Mart{\'{\i}}n},
  {Feldt}, {Henning}, {Pott}, {Bonnefoy}, {Bouwman}, {Lacour}, {Mueller},
  {Juh{\'a}sz}, {Crida}, {Chauvin}, {Andrews}, {Wilner}, {Kraus}, {Dahm},
  {Robitaille}, {Jang-Condell}, {Abe}, {Akiyama}, {Brandner}, {Brandt},
  {Carson}, {Egner}, {Follette}, {Goto}, {Guyon}, {Hayano}, {Hayashi},
  {Hayashi}, {Hodapp}, {Ishii}, {Iye}, {Janson}, {Kandori}, {Knapp}, {Kudo},
  {Kusakabe}, {Kuzuhara}, {Mayama}, {McElwain}, {Matsuo}, {Miyama}, {Morino},
  {Nishimura}, {Pyo}, {Serabyn}, {Suto}, {Suzuki}, {Takami}, {Takato},
  {Terada}, {Tomono}, {Turner}, {Watanabe}, {Yamada}, {Takami}, {Usuda}, \&
  {Tamura}}]{grady13}
{Grady}, C.~A., {et~al.} 2013, \apj, 762, 48

\bibitem[{{Hashimoto} {et~al.}(2012){Hashimoto}, {Dong}, {Kudo}, {Honda},
  {McClure}, {Zhu}, {Muto}, {Wisniewski}, {Abe}, {Brandner}, {Brandt},
  {Carson}, {Egner}, {Feldt}, {Fukagawa}, {Goto}, {Grady}, {Guyon}, {Hayano},
  {Hayashi}, {Hayashi}, {Henning}, {Hodapp}, {Ishii}, {Iye}, {Janson},
  {Kandori}, {Knapp}, {Kusakabe}, {Kuzuhara}, {Kwon}, {Matsuo}, {Mayama},
  {McElwain}, {Miyama}, {Morino}, {Moro-Martin}, {Nishimura}, {Pyo}, {Serabyn},
  {Suenaga}, {Suto}, {Suzuki}, {Takahashi}, {Takami}, {Takato}, {Terada},
  {Thalmann}, {Tomono}, {Turner}, {Watanabe}, {Yamada}, {Takami}, {Usuda}, \&
  {Tamura}}]{hashimoto12}
{Hashimoto}, J., {et~al.} 2012, \apjl, 758, L19

\bibitem[{{Hashimoto} {et~al.}(2011){Hashimoto}, {Tamura}, {Muto}, {Kudo},
  {Fukagawa}, {Fukue}, {Goto}, {Grady}, {Henning}, {Hodapp}, {Honda},
  {Inutsuka}, {Kokubo}, {Knapp}, {McElwain}, {Momose}, {Ohashi}, {Okamoto},
  {Takami}, {Turner}, {Wisniewski}, {Janson}, {Abe}, {Brandner}, {Carson},
  {Egner}, {Feldt}, {Golota}, {Guyon}, {Hayano}, {Hayashi}, {Hayashi}, {Ishii},
  {Kandori}, {Kusakabe}, {Matsuo}, {Mayama}, {Miyama}, {Morino}, {Moro-Martin},
  {Nishimura}, {Pyo}, {Suto}, {Suzuki}, {Takato}, {Terada}, {Thalmann},
  {Tomono}, {Watanabe}, {Yamada}, {Takami}, \& {Usuda}}]{hashimoto11}
---. 2011, \apjl, 729, L17

\bibitem[{{Jang-Condell} \& {Turner}(2013)}]{jangcondell13}
{Jang-Condell}, H., \& {Turner}, N.~J. 2013, \apj, 772, 34

\bibitem[{{Janson} {et~al.}(2016){Janson}, {Thalmann}, {Boccaletti}, {Maire},
  {Zurlo}, {Marzari}, {Meyer}, {Carson}, {Augereau}, {Garufi}, {Henning},
  {Desidera}, {Asensio-Torres}, \& {Pohl}}]{janson16}
{Janson}, M., {et~al.} 2016, \apjl, 816, L1

\bibitem[{{Jovanovic} {et~al.}(2015){Jovanovic}, {Martinache}, {Guyon},
  {Clergeon}, {Singh}, {Kudo}, {Garrel}, {Newman}, {Doughty}, {Lozi}, {Males},
  {Minowa}, {Hayano}, {Takato}, {Morino}, {Kuhn}, {Serabyn}, {Norris},
  {Tuthill}, {Schworer}, {Stewart}, {Close}, {Huby}, {Perrin}, {Lacour},
  {Gauchet}, {Vievard}, {Murakami}, {Oshiyama}, {Baba}, {Matsuo}, {Nishikawa},
  {Tamura}, {Lai}, {Marchis}, {Duchene}, {Kotani}, \& {Woillez}}]{jovanovic15}
{Jovanovic}, N., {et~al.} 2015, \pasp, 127, 890

\bibitem[{{Kim} {et~al.}(1994){Kim}, {Martin}, \& {Hendry}}]{kim94}
{Kim}, S.-H., {Martin}, P.~G., \& {Hendry}, P.~D. 1994, \apj, 422, 164

\bibitem[{{Kley} \& {Nelson}(2012)}]{kley12}
{Kley}, W., \& {Nelson}, R.~P. 2012, \araa, 50, 211

\bibitem[{{Konishi} {et~al.}(2016){Konishi}, {Grady}, {Schneider}, {Shibai},
  {McElwain}, {Nesvold}, {Kuchner}, {Carson}, {Debes}, {Gaspar}, {Henning},
  {Hines}, {Hinz}, {Jang-Condell}, {Moro-Martin}, {Perrin}, {Rodigas},
  {Serabyn}, {Silverstone}, {Stark}, {Tamura}, {Weinberger}, \&
  {Wisniewski}}]{konishi16}
{Konishi}, M., {et~al.} 2016, ArXiv e-prints

\bibitem[{{Kusakabe} {et~al.}(2012){Kusakabe}, {Grady}, {Sitko}, {Hashimoto},
  {Kudo}, {Fukagawa}, {Muto}, {Wisniewski}, {Min}, {Mayama}, {Werren}, {Day},
  {Beerman}, {Lynch}, {Russell}, {Brafford}, {Kuzuhara}, {Brandt}, {Abe},
  {Brandner}, {Carson}, {Egner}, {Feldt}, {Goto}, {Guyon}, {Hayano}, {Hayashi},
  {Hayashi}, {Henning}, {Hodapp}, {Ishii}, {Iye}, {Janson}, {Kandori}, {Knapp},
  {Matsuo}, {McElwain}, {Miyama}, {Morino}, {Moro-Martin}, {Nishimura}, {Pyo},
  {Suto}, {Suzuki}, {Takami}, {Takato}, {Terada}, {Thalmann}, {Tomono},
  {Turner}, {Watanabe}, {Yamada}, {Takami}, {Usuda}, \& {Tamura}}]{kusakabe12}
{Kusakabe}, N., {et~al.} 2012, \apj, 753, 153

\bibitem[{{Lenzen} {et~al.}(2003){Lenzen}, {Hartung}, {Brandner}, {Finger},
  {Hubin}, {Lacombe}, {Lagrange}, {Lehnert}, {Moorwood}, \&
  {Mouillet}}]{lenzen03}
{Lenzen}, R., {et~al.} 2003, in \procspie, Vol. 4841, Instrument Design and
  Performance for Optical/Infrared Ground-based Telescopes, ed. M.~{Iye} \&
  A.~F.~M. {Moorwood}, 944--952

\bibitem[{{Lovelace} {et~al.}(1999){Lovelace}, {Li}, {Colgate}, \&
  {Nelson}}]{lovelace99}
{Lovelace}, R.~V.~E., {Li}, H., {Colgate}, S.~A., \& {Nelson}, A.~F. 1999,
  \apj, 513, 805

\bibitem[{{Lucy}(1999)}]{lucy99}
{Lucy}, L.~B. 1999, \aap, 344, 282

\bibitem[{{Macintosh} {et~al.}(2008){Macintosh}, {Graham}, {Palmer}, {Doyon},
  {Dunn}, {Gavel}, {Larkin}, {Oppenheimer}, {Saddlemyer}, {Sivaramakrishnan},
  {Wallace}, {Bauman}, {Erickson}, {Marois}, {Poyneer}, \&
  {Soummer}}]{macintosh08}
{Macintosh}, B.~A., {et~al.} 2008, in Society of Photo-Optical Instrumentation
  Engineers (SPIE) Conference Series, Vol. 7015, Society of Photo-Optical
  Instrumentation Engineers (SPIE) Conference Series

\bibitem[{{Mayama} {et~al.}(2012){Mayama}, {Hashimoto}, {Muto}, {Tsukagoshi},
  {Kusakabe}, {Kuzuhara}, {Takahashi}, {Kudo}, {Dong}, {Fukagawa}, {Takami},
  {Momose}, {Wisniewski}, {Follette}, {Abe}, {Akiyama}, {Brandner}, {Brandt},
  {Carson}, {Egner}, {Feldt}, {Goto}, {Grady}, {Guyon}, {Hayano}, {Hayashi},
  {Hayashi}, {Henning}, {Hodapp}, {Ishii}, {Iye}, {Janson}, {Kandori}, {Kwon},
  {Knapp}, {Matsuo}, {McElwain}, {Miyama}, {Morino}, {Moro-Martin},
  {Nishimura}, {Pyo}, {Serabyn}, {Suto}, {Suzuki}, {Takato}, {Terada},
  {Thalmann}, {Tomono}, {Turner}, {Watanabe}, {Yamada}, {Takami}, {Usuda}, \&
  {Tamura}}]{mayama12}
{Mayama}, S., {et~al.} 2012, \apjl, 760, L26

\bibitem[{{Muto} {et~al.}(2012){Muto}, {Grady}, {Hashimoto}, {Fukagawa},
  {Hornbeck}, {Sitko}, {Russell}, {Werren}, {Cur{\'e}}, {Currie}, {Ohashi},
  {Okamoto}, {Momose}, {Honda}, {Inutsuka}, {Takeuchi}, {Dong}, {Abe},
  {Brandner}, {Brandt}, {Carson}, {Egner}, {Feldt}, {Fukue}, {Goto}, {Guyon},
  {Hayano}, {Hayashi}, {Hayashi}, {Henning}, {Hodapp}, {Ishii}, {Iye},
  {Janson}, {Kandori}, {Knapp}, {Kudo}, {Kusakabe}, {Kuzuhara}, {Matsuo},
  {Mayama}, {McElwain}, {Miyama}, {Morino}, {Moro-Martin}, {Nishimura}, {Pyo},
  {Serabyn}, {Suto}, {Suzuki}, {Takami}, {Takato}, {Terada}, {Thalmann},
  {Tomono}, {Turner}, {Watanabe}, {Wisniewski}, {Yamada}, {Takami}, {Usuda}, \&
  {Tamura}}]{muto12}
{Muto}, T., {et~al.} 2012, \apjl, 748, L22

\bibitem[{{P{\'e}ricaud} {et~al.}(2014){P{\'e}ricaud}, {Di Folco}, {Dutrey},
  {Augereau}, {Pi{\'e}tu}, \& {Guilloteau}}]{pericaud14}
{P{\'e}ricaud}, J., {Di Folco}, E., {Dutrey}, A., {Augereau}, J.-C.,
  {Pi{\'e}tu}, V., \& {Guilloteau}, S. 2014, in Thirty years of Beta Pic and
  Debris Disks Studies

\bibitem[{{Pinilla} {et~al.}(2015{\natexlab{a}}){Pinilla}, {de Boer},
  {Benisty}, {Juh{\'a}sz}, {de Juan Ovelar}, {Dominik}, {Avenhaus},
  {Birnstiel}, {Girard}, {Huelamo}, {Isella}, \& {Milli}}]{pinilla15j1604}
{Pinilla}, P., {et~al.} 2015{\natexlab{a}}, \aap, 584, L4

\bibitem[{{Pinilla} {et~al.}(2015{\natexlab{b}}){Pinilla}, {de Juan Ovelar},
  {Ataiee}, {Benisty}, {Birnstiel}, {van Dishoeck}, \&
  {Min}}]{pinilla15twoplanets}
{Pinilla}, P., {de Juan Ovelar}, M., {Ataiee}, S., {Benisty}, M., {Birnstiel},
  T., {van Dishoeck}, E.~F., \& {Min}, M. 2015{\natexlab{b}}, \aap, 573, A9

\bibitem[{{Pohl} {et~al.}(2015){Pohl}, {Pinilla}, {Benisty}, {Ataiee},
  {Juh{\'a}sz}, {Dullemond}, {Van Boekel}, \& {Henning}}]{pohl15}
{Pohl}, A., {Pinilla}, P., {Benisty}, M., {Ataiee}, S., {Juh{\'a}sz}, A.,
  {Dullemond}, C.~P., {Van Boekel}, R., \& {Henning}, T. 2015, \mnras, 453,
  1768

\bibitem[{{Quanz} {et~al.}(2013){Quanz}, {Avenhaus}, {Buenzli}, {Garufi},
  {Schmid}, \& {Wolf}}]{quanz13gap}
{Quanz}, S.~P., {Avenhaus}, H., {Buenzli}, E., {Garufi}, A., {Schmid}, H.~M.,
  \& {Wolf}, S. 2013, \apjl, 766, L2

\bibitem[{{Rapson} {et~al.}(2015){Rapson}, {Kastner}, {Millar-Blanchaer}, \&
  {Dong}}]{rapson15twhya}
{Rapson}, V.~A., {Kastner}, J.~H., {Millar-Blanchaer}, M.~A., \& {Dong}, R.
  2015, \apjl, 815, L26

\bibitem[{{Schmid} {et~al.}(2006){Schmid}, {Joos}, \& {Tschan}}]{schmid06}
{Schmid}, H.~M., {Joos}, F., \& {Tschan}, D. 2006, \aap, 452, 657

\bibitem[{{Shakura} \& {Sunyaev}(1973)}]{shakura73}
{Shakura}, N.~I., \& {Sunyaev}, R.~A. 1973, \aap, 24, 337

\bibitem[{{Stapelfeldt} {et~al.}(1999){Stapelfeldt}, {Watson}, {Krist},
  {Burrows}, {Crisp}, {Ballester}, {Clarke}, {Evans}, {Gallagher}, {Griffiths},
  {Hester}, {Hoessel}, {Holtzman}, {Mould}, {Scowen}, \&
  {Trauger}}]{stapelfeldt99}
{Stapelfeldt}, K.~R., {et~al.} 1999, \apjl, 516, L95

\bibitem[{{Stolker} {et~al.}(2016){Stolker}, {Dominik}, {Avenhaus}, {Min}, {de
  Boer}, {Ginski}, {Schmid}, {Juhasz}, {Bazzon}, {Waters}, {Garufi},
  {Augereau}, {Benisty}, {Boccaletti}, {Henning}, {Maire}, {Menard}, {Meyer},
  {Langlois}, {Pinte}, {Quanz}, {Thalmann}, {Beuzit}, {Carbillet}, {Costille},
  {Dohlen}, {Feldt}, {Gisler}, {Mouillet}, {Pavlov}, {Perret}, {Petit},
  {Pragt}, {Rochat}, {Roelfsema}, {Salasnich}, {Soenke}, \&
  {Wildi}}]{stolker16}
{Stolker}, T., {et~al.} 2016, ArXiv e-prints

\bibitem[{{Tamura} {et~al.}(2006){Tamura}, {Hodapp}, {Takami}, {Abe}, {Suto},
  {Guyon}, {Jacobson}, {Kandori}, {Morino}, {Murakami}, {Stahlberger},
  {Suzuki}, {Tavrov}, {Yamada}, {Nishikawa}, {Ukita}, {Hashimoto}, {Izumiura},
  {Hayashi}, {Nakajima}, \& {Nishimura}}]{tamura06}
{Tamura}, M., {et~al.} 2006, in \procspie, Vol. 6269, Society of Photo-Optical
  Instrumentation Engineers (SPIE) Conference Series, 62690V

\bibitem[{{Tang} {et~al.}(2012){Tang}, {Guilloteau}, {Pi{\'e}tu}, {Dutrey},
  {Ohashi}, \& {Ho}}]{tang12}
{Tang}, Y.-W., {Guilloteau}, S., {Pi{\'e}tu}, V., {Dutrey}, A., {Ohashi}, N.,
  \& {Ho}, P.~T.~P. 2012, \aap, 547, A84

\bibitem[{{Thalmann} {et~al.}(2015){Thalmann}, {Mulders}, {Janson}, {Olofsson},
  {Benisty}, {Avenhaus}, {Quanz}, {Schmid}, {Henning}, {Buenzli}, {M{\'e}nard},
  {Carson}, {Garufi}, {Messina}, {Dominik}, {Leisenring}, {Chauvin}, \&
  {Meyer}}]{thalmann15}
{Thalmann}, C., {et~al.} 2015, \apjl, 808, L41

\bibitem[{{Wagner} {et~al.}(2015){Wagner}, {Apai}, {Kasper}, \&
  {Robberto}}]{wagner15100453}
{Wagner}, K., {Apai}, D., {Kasper}, M., \& {Robberto}, M. 2015, \apjl, 813, L2

\bibitem[{{Wang} {et~al.}(2015){Wang}, {Graham}, {Pueyo}, {Nielsen},
  {Millar-Blanchaer}, {De Rosa}, {Kalas}, {Ammons}, {Bulger}, {Cardwell},
  {Chen}, {Chiang}, {Chilcote}, {Doyon}, {Draper}, {Duch{\^e}ne}, {Esposito},
  {Fitzgerald}, {Goodsell}, {Greenbaum}, {Hartung}, {Hibon}, {Hinkley}, {Hung},
  {Ingraham}, {Larkin}, {Macintosh}, {Maire}, {Marchis}, {Marois}, {Matthews},
  {Morzinski}, {Oppenheimer}, {Patience}, {Perrin}, {Rajan}, {Rantakyr{\"o}},
  {Sadakuni}, {Serio}, {Sivaramakrishnan}, {Soummer}, {Thomas}, {Ward-Duong},
  {Wiktorowicz}, \& {Wolff}}]{wang15}
{Wang}, J.~J., {et~al.} 2015, \apjl, 811, L19

\bibitem[{{Weinberger} {et~al.}(2000){Weinberger}, {Rich}, {Becklin},
  {Zuckerman}, \& {Matthews}}]{weinberger00}
{Weinberger}, A.~J., {Rich}, R.~M., {Becklin}, E.~E., {Zuckerman}, B., \&
  {Matthews}, K. 2000, \apj, 544, 937

\bibitem[{{Whitney} {et~al.}(2013){Whitney}, {Robitaille}, {Bjorkman}, {Dong},
  {Wolff}, {Wood}, \& {Honor}}]{whitney13}
{Whitney}, B.~A., {Robitaille}, T.~P., {Bjorkman}, J.~E., {Dong}, R., {Wolff},
  M.~J., {Wood}, K., \& {Honor}, J. 2013, \apjs, 207, 30

\bibitem[{{Zhu} {et~al.}(2015){Zhu}, {Dong}, {Stone}, \&
  {Rafikov}}]{zhu15densitywaves}
{Zhu}, Z., {Dong}, R., {Stone}, J.~M., \& {Rafikov}, R.~R. 2015, \apj, 813, 88

\bibitem[{{Zhu} {et~al.}(2011){Zhu}, {Nelson}, {Hartmann}, {Espaillat}, \&
  {Calvet}}]{zhu11}
{Zhu}, Z., {Nelson}, R.~P., {Hartmann}, L., {Espaillat}, C., \& {Calvet}, N.
  2011, \apj, 729, 47

\end{thebibliography}
\end{document}